\documentclass{ieeeaccess}
\usepackage{cite}
\usepackage{amsmath,amssymb,amsfonts}
\usepackage{algorithmic}
\usepackage{graphicx}

\usepackage{textcomp}
\usepackage{hyperref}
\usepackage{amsmath}
\usepackage{amsmath,amssymb,amsfonts}
\usepackage{algorithmic}
\usepackage{algorithm}
\usepackage{lscape}
\usepackage{multirow}
\usepackage{subcaption}
\def\orcid#1{\kern .08em\href{https://orcid.org/#1}{\includegraphics[keepaspectratio,width=0.7em]{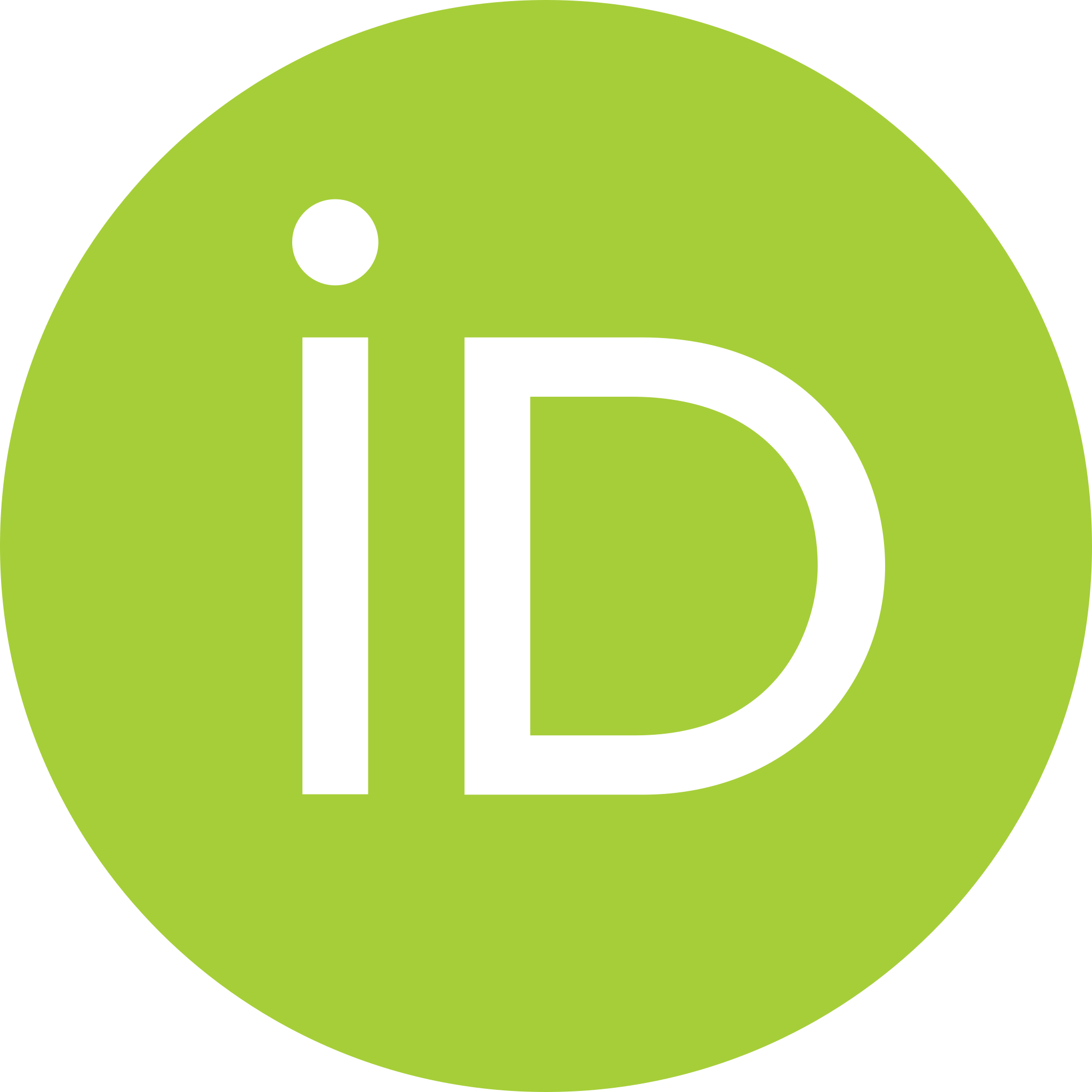}}}
\def\BibTeX{{\rm B\kern-.05em{\sc i\kern-.025em b}\kern-.08em
    T\kern-.1667em\lower.7ex\hbox{E}\kern-.125emX}}
\begin{document}
\history{Date of publication xxxx 00, 0000, date of current version xxxx 00, 0000.}
\doi{10.1109/ACCESS.2017.DOI}

\title{DSLR-CNN: Efficient CNN Acceleration using Digit-Serial Left-to-Right Arithmetic}

\author{\uppercase{Malik Zohaib Nisar}\href{https://orcid.org/0009-0009-9164-3186}{\includegraphics[scale=0.07]{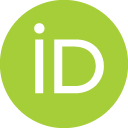}}\authorrefmark{1},
\uppercase{Muhammad Sohail Ibrahim}\href{https://orcid.org/0000-0002-1387-0879}{\includegraphics[scale=0.07]{ORCIDiD_icon128x128.png}}\authorrefmark{1},
\uppercase{Saeid Gorgin}\href{https://orcid.org/0000-0001-5898-4872}{\includegraphics[scale=0.07]{ORCIDiD_icon128x128.png}}\authorrefmark{1,3}
\uppercase{Muhammad Usman}\href{https://orcid.org/0000-0002-3393-5211}{\includegraphics[scale=0.07]{ORCIDiD_icon128x128.png}}\authorrefmark{1,2} and \uppercase{Jeong-A Lee}\href{https://orcid.org/0000-0002-5166-0629}{\includegraphics[scale=0.07]{ORCIDiD_icon128x128.png}}\authorrefmark{1}}


\address[1]{ Department of Computer Engineering, College of IT Convergence, Chosun University, Gwangju, 61452, Republic of Korea
(Email: \{zohaib, msohail, gorgin, usman, jalee\}@chosun.ac.kr).}
\address[2]{Faculty of Informatics and Data Science, University of Regensburg, Regensburg, 93053, Germany (Email: muhammad.usman@ur.de).}
\address[3]{Department of Electrical Engineering and Information Technology, Iranian Research Organization for Science and Technology (IROST), Tehran 33535-111, Iran.}

\tfootnote{This research was supported by the Basic Science Research Program funded by the Ministry of Education through the National Research Foundation of Korea  $(NRF-2020R1I1A3063857)$. The EDA tool was supported by the IC Design Education Center (IDEC), Korea.}

\markboth
{Author \headeretal: Preparation of Papers for IEEE TRANSACTIONS and JOURNALS}
{Author \headeretal: Preparation of Papers for IEEE TRANSACTIONS and JOURNALS}

\corresp{Corresponding author(s): Muhammad Usman, Jeong-A Lee (e-mail: {usman,jalee}@chosun.ac.kr).}

\begin{abstract}
Digit-serial arithmetic has emerged as a viable approach for designing hardware accelerators, reducing interconnections, area utilization, and power consumption. However, conventional methods suffer from performance and latency issues. To address these challenges, we propose an accelerator design using left-to-right (LR) arithmetic, which performs computations in a most-significant digit first (MSDF) manner, enabling digit-level pipelining. This leads to substantial performance improvements and reduced latency. The processing engine is designed for convolutional neural networks (CNNs), which includes low-latency LR multipliers and adders for digit-level parallelism. The proposed DSLR-CNN is implemented in Verilog and synthesized with Synopsys design compiler using GSCL 45nm technology, the DSLR-CNN accelerator was evaluated on AlexNet, VGG-16, and ResNet-18 networks. Results show significant improvements across key performance evaluation metrics, including response time, peak performance, power consumption, operational intensity, area efficiency, and energy efficiency. The peak performance measured in GOPS of the proposed design is 4.37$\times$ to 569.11$\times$ higher than contemporary designs, and it achieved 3.58$\times$ to 44.75$\times$ higher peak energy efficiency (TOPS/W), outperforming conventional bit-serial designs.
\end{abstract}


\titlepgskip=-15pt

\maketitle
\section{Introduction}
\label{sec:introduction}
\PARstart{I}{n} modern computing systems, deep neural networks (DNNs) have become indispensable in various artificial intelligence subfields, such as computer vision \cite{al2023automated}, object identification \cite{ye2023real}, and speech recognition \cite{mahmoud2023comparative}. CNNs, in particular, play a crucial role in managing complex tasks within these domains. They are essential in applications like bioinformatics \cite{usman2021aop}, modulation classification \cite{usman2020amc,nisar2023lightweight}, deep symbolic optimization \cite{usama2022data} and the development of hardware systems \cite{dean20201}. CNNs have achieved remarkable performance, often surpassing human capabilities. However, the high computational complexity required by CNNs poses significant challenges for both computational performance and energy efficiency.
While graphics processing units (GPUs) can address the high computational demands of CNNs, they also consume substantial energy. The performance and complexity of CNNs are well-documented, with recent research indicating that the number of layers significantly impacting overall performance \cite{kwon2019understanding}. Generally, adding more layers enhances the networks ability to extract intricate features, thus improving performance. However, deeper networks necessitate more parameters, increasing the computational and memory demands for effective training and inference. As the demand for reduced computation, memory footprint, and high throughput for CNN inference grows, domain-specific hardware accelerators have become increasingly vital \cite{dally1998digital, hennessy2019new}. The majority of CNN operations occur in the computation of convolutional and fully connected layers, accounting for approximately $90\%$ of multiply-and-accumulate (MAC) operations \cite{jain2018compensated}. These MAC operations significantly influence the computational costs, latency, and performance efficiency of the network. Consequently, researchers are exploring specialized digit-serial or bit-serial computation units to achieve efficient computation and communication in CNNs.

Unlike the conventional bit-parallel technique, where multiple bits are processed simultaneously in a single clock cycle, bit-serial computation processes each bit of a number sequentially, requiring multiple cycles to complete the operation. The sequential processing reduces the size of required storage and arithmetic logic units, making it suitable for hardware with limited resources or where cost-effectiveness is crucial. Furthermore, its streamlined computational circuitry leads to lower power consumption, a benefit for battery-powered mobile and embedded devices, where power efficiency is a one of the primary challenges. To show the effectiveness of bit-serial arithmetic, various CNN accelerator designs have been proposed to enhance the efficiency and performance of CNN, such as exploiting the potential for data reuse through the use of accelerators \cite{chen2016eyeriss}, taking advantage of the high number of zeroes in weight and activation matrices (known as sparsity exploitation) \cite{liu2023cnn, chen2014diannao}. Additionally, variable bit-width architectures have been used to reduce bit-width without sacrificing accuracy \cite{judd2016stripes, sun2024cim2pq, sharma2018bit, sharify2018loom}. These digit-serial techniques efficiently manage the data flow, can achieve faster inference times, and improve throughput in CNN computations. 
 This method offers compelling advantages: 1) it drastically reduces the need for hardware resources \cite{cheng2024leveraging}, 2) offers exceptional energy efficiency and flexibility for implementing neural network accelerators in low-power and energy-constrained systems \cite{zhao2022network}, 3) requires fewer hardware resources requirements compared to parallel processing, 4) provides opportunity for dynamic precision adjustment at runtime, enabling significant power savings by adjusting the network precision according to the specific application and datasets \cite{metz2023embedded}, 5) provides opportunity to exploit sparsity \cite{li2023precision}. 
  However, despite its benefits, bit-serial computation faces challenges. The sequential processing can potentially introduce computational delays, raising concerns about its impact on overall latency and computational performance. Therefore, developing a suitable strategy becomes crucial to ensure that any compromise in computational delays and performance remains within acceptable limits while minimizing data width \cite{cheng2024leveraging}. Existing bit-serial strategies typically perform computations in a conventional right-to-left manner, where each arithmetic unit waits for the completion of the preceding operation to start its computation. This introduces idle time for the subsequent units, creating a bottleneck, reducing the overall processing speed, and limiting the performance and scalability of the architecture, especially in tasks requiring high throughput and low latency, such as real-time processing or high-performance computing applications \cite{judd2016stripes, sharify2018loom}. 
  
  To address these limitations, we propose a CNN accelerator that utilizes an unconventional LR computation pattern known as online or LR arithmetic \cite{ercegovac2004digital}. The superiority of this method is shown in the acceleration of several machine learning algorithms, such as kNN-MSDF \cite{gorgin2022energy} and K-means \cite{gorgin2022knn}. In these approaches, computations are executed serially, digit by digit, in a MSDF order. This means the input is provided from left to right, and the output of the most significant digit is generated first. The first digit of the result is produced after a fixed small delay known as the online delay ($\delta$), during which a few input digits are processed. The adoption of LR arithmetic reduces latency, increases performance, and improves energy efficiency by minimizing interconnects and reducing memory footprint, making it an efficient choice for inference on resource-constrained devices \cite{ibrahim2023dslot, ibrahim2024echo}. The main contributions to our research can be categorized as follows:

\begin{itemize}
    \item A low-latency, high-throughput convolution SoP unit has been designed using LR arithmetic for convolution computations in CNNs, with the aim of reducing response time, increasing performance, and enhancing energy efficiency.
     \item To demonstrate the effectiveness of the proposed DSLR-CNN, we examine the proposed strategy on the convolution layers of the AlexNet, VGG-16, and ResNet-18 networks.
    \item The proposed design implemented on RTL (Verilog) and synthesized using the Synopsys design compiler with GSCL $45$nm technology. It is evaluated and compared with the conventional bit-serial baseline design in terms of duration, performance, area utilization, power consumption, and energy efficiency. 
    \item Finally, the overall performance and energy efficiency of the DSLR-CNN are compared with various state-of-the-art accelerators. 
   
\end{itemize}

The rest of the paper is organized as follows. Section \ref{sec:background} presents the background of the study. Section \ref{sec: DSLR-CNN} presents the details of the proposed DSLR-CNN design. The results and discussion of the proposed methodology are presented in Section \ref{sec: Results}. Related work is provided in Section \ref{sec:RW} followed by the conclusion in Section \ref{sec: conclusion}.

\section{Background}\label{sec:background}
This section mainly focuses on the introduction of CNN and its function. Then, we explore the utilization of LR arithmetic techniques within DNNs. Specifically, we detail the design considerations and implementation strategies for MSDF multipliers and adders.

\subsection{Convolution Neural network}
\begin{figure}
    \centering
    \includegraphics[width=\linewidth]{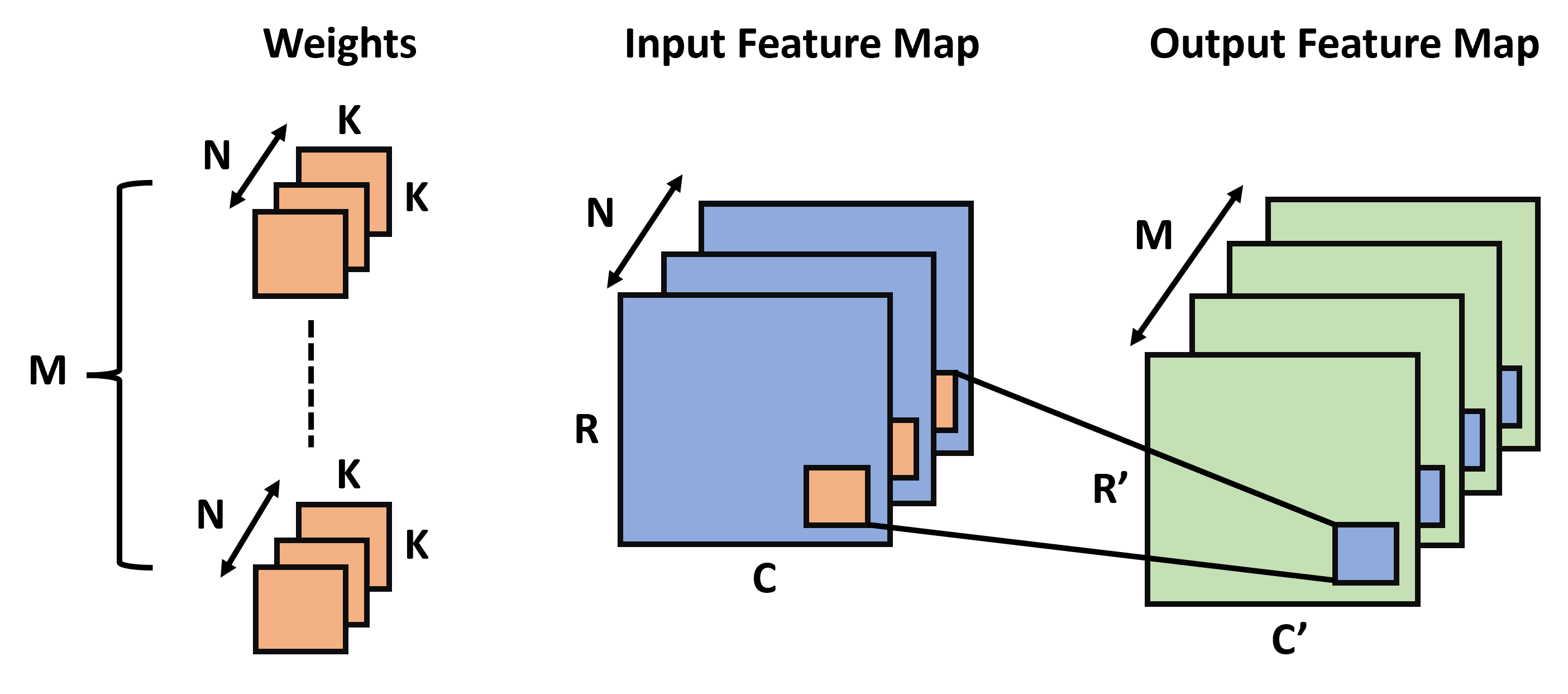}
    \caption{Illustration of a Convolution Operation.}
    \label{fig: cnn}
\end{figure}

CNNs are a highly effective type of machine learning model renowned for their versatility, particularly in the field of image recognition. CNNs typically comprise a series of convolutional layers followed by one or more fully connected (dense) neural network layers for classification purposes. Fig.~\ref{fig: cnn} illustrates a single convolutional layer to provide a visual representation. In this context, let us consider the input to a convolutional layer consisting of input feature maps with $N$ channels, each consisting of $R$ $\times$ $C$ values. The layer employs $M$ sets of $N$ $\times$ $K$ $\times$ $K$ filters, with their weights determined via a learning algorithm such as backpropagation. Each filter set slides across the input feature map with a stride of $S$, multiplying its values with the corresponding values from the input feature maps at each position. The resulting products are added to produce one value in an output feature map. This process is repeated for each of the $M$ filter sets, generating $M$ output feature maps of dimensions $R'$ $\times$ $C'$, where $R'$ = $(R - K + 2P)/S + 1$ and $C' = (C - K + 2P)/S + 1$. Subsequently, a bias value (also determined via backpropagation) is added to each of the $M$ output feature maps. Finally, the resulting feature maps undergo a non-linear activation operation, such as the Rectified Linear Unit (ReLU), and optionally a subsampling operation, such as maxpooling. Table~\ref{tab:conf} shows the nomenclature related to the CNN and its tiling parameters.

\renewcommand{\arraystretch}{1.6}
\begin{table}[!ht]\label{parameters}
\centering
\caption{Nomenclature}\label{tab:conf}
\begin{tabular}{c|l} \hline \hline
$N$& Input Feature Map  \\ \hline
$M$ & Output Feature Map \\ \hline
$R$& Feature Map Row \\ \hline 
$C$ & Feature Map Column \\ \hline
$K$ & Kernel \\ \hline
$T_{n}$ & Input Tiling Factor \\ \hline
$T_{m}$ & Output Tiling Factor \\ \hline
$T_{r}$ & Row Tiling Factor \\ \hline
$T_{c}$ & Column Tiling Factor\\ \hline\hline
\end{tabular}
\end{table}

\subsection{Left-to-Right Arithmetic}
\begin{figure}[!ht]
	\begin{center}
 		\includegraphics*[width=7cm]{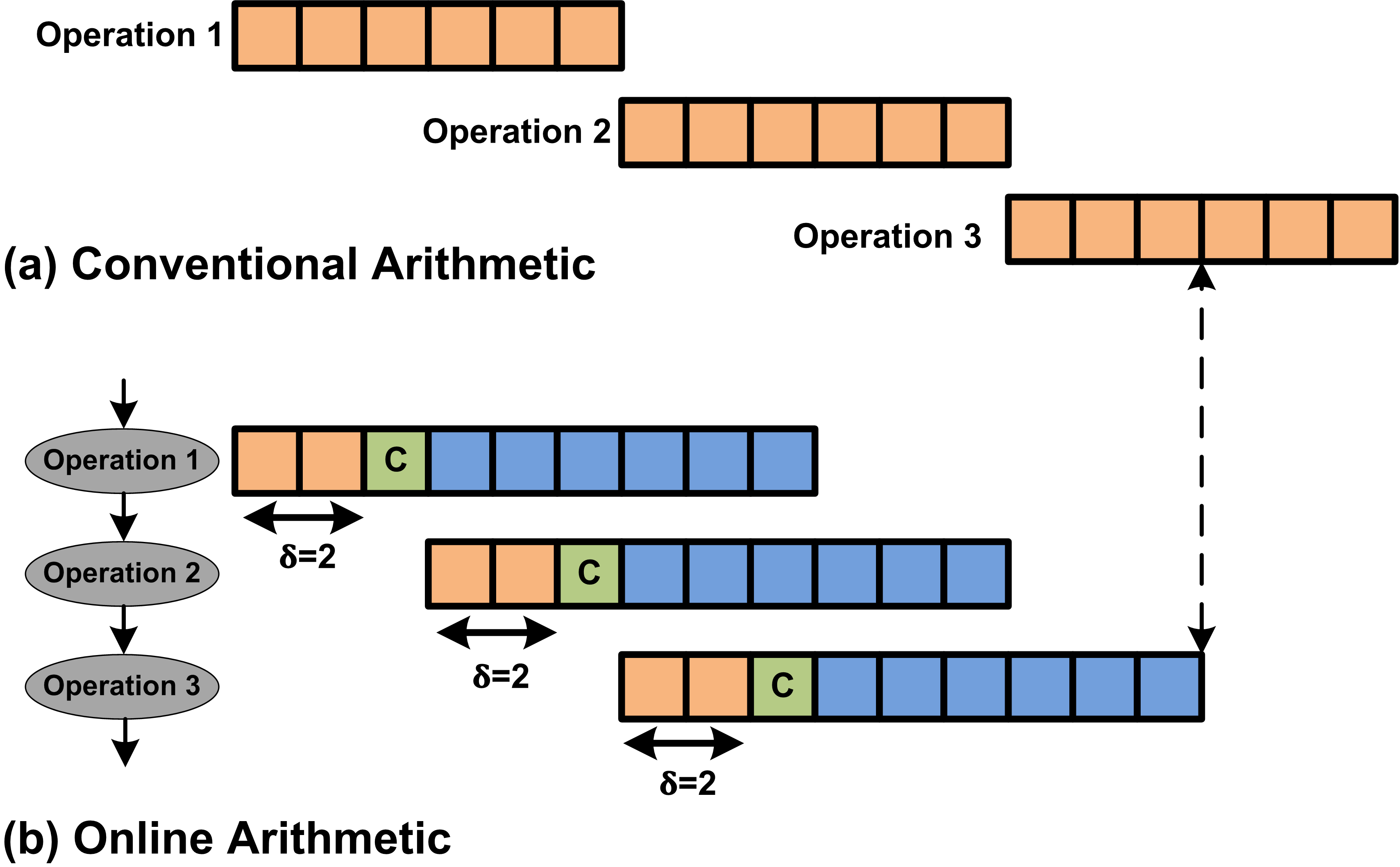}
 	\end{center}
 	\caption{Comparative Timing Analysis of Conventional Arithmetic vs. Online Arithmetic for Sequential Interdependent Operations.}
 	\label{fig: OArith}
 \end{figure}

LR arithmetic is an unconventional arithmetic paradigm where inputs are fed and results are computed digit-serially in a MSDF fashion, subsequently minimizing interconnect lines, necessitating low-bandwidth communication, simplifying the interface, and enhancing energy efficiency \cite{ercegovac2004digital}, \cite{joseph2018algorithms}. The generation of the most significant output digit on the basis of a few input digits is enabled by the use of a redundant number system, which results in a slight increase in the overall area of the computation units (online multipliers, adders, etc.). To generate the first digit of the result, \(\delta \) digits from the input operands are needed, which when processed, the MSD of the result is generated. Subsequently, an additional digit in the result is obtained for each additional digit in the input operands. The result of the current operation can be directly fed to the input of the succeeding operation, achieving digit-level pipelining. Fig.~\ref{fig: OArith} illustrates the computation patterns of LR and conventional arithmetic with  \(\delta \) = $2$ and a compute cycle of c = $1$. 

The key timing difference between these methods lies in their operation sequences. Conventional arithmetic operators must wait for the entire previous computation to complete, as illustrated in Fig.~\ref{fig: OArith}(a). However, in LR arithmetic, computations can begin as soon as the MSD of the result is available from the preceding operation, which occurs after  \(\delta \) + compute cycles, while the remaining inputs are processed sequentially, as depicted in Fig.~\ref{fig: OArith}(b). This sequential processing allows dependent operations to be executed almost simultaneously, enhancing efficiency. 

As stated above, in order to compute the output using only partial information from the inputs, it is necessary to employ the redundant number system \cite{ercegovac2004digital}. A radix-$r$ weighted number system is considered redundant if its digit set is also redundant, allowing multiple representations of the same value. This flexibility enables the selection of an output digit at any given computation stage \cite{ ercegovac2020reducing, usman2021multiplier}. In this work, we utilize signed-digit (SD) representation on a redundant digit set of \{$-1, 0, 1$\} for all online operators. With SD representation, we denote the $j\textsuperscript{th}$ digit of a number as $x_j$. The numerical value of $x_j$ is given by $(x^+_j, x^-_j)$, where $x^+_j$ and $x^-_j$ are single bits. The inputs and outputs are represented by \eqref{eq1}.

\begin{equation}\label{eq1}
x[j] = \sum _{i=1}^{j+\delta}x_{i}r^{-i},
\end{equation}
where the square bracket indicates the iteration, while the subscripts indicate the index of each digit. 

The design of multipliers has shown a significant effect on the performance of signal processing and machine learning applications in terms of power and area \cite{malathi2024fpga}, \cite{tang2023area}, \cite{reddy2022low}. For partial product generation and reduction, traditional multipliers can be divided into linear array multipliers and tree multipliers. The digit parallel computation utilized by the multipliers requires full bandwidth interconnection, accounting for increased latency, power, and energy requirements. 

Herein, we propose to use the LR arithmetic-based multiplier to perform multiplication in convolution. Since the weights are readily available during the inference of CNNs, we opt to use the online multiplier with one operand in a serial MSDF manner, while the other operand as constant. The input image is processed serially while the weight is employed in parallel. The constant weight is shown in \eqref{eq2}.

\begin{equation}\label{eq2}
    Y[j] = Y = -y_{0} \cdot r^{0} + \sum_{i=1}^{n} y_{i} r^{-i}
\end{equation}

\begin{figure}[ht]
	\begin{center}
 		\includegraphics*[width=8.5cm]{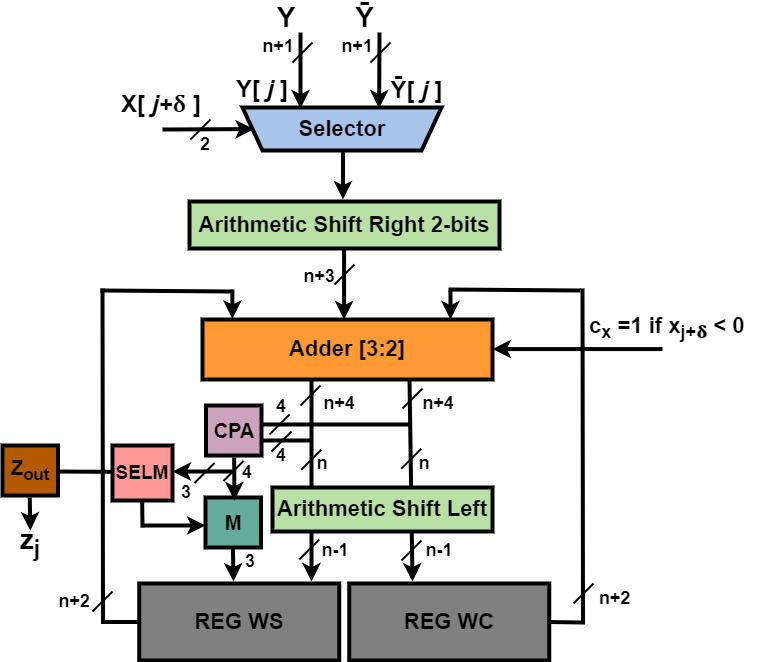}
 	\end{center}
 	\caption{LR Serial-Parallel Multiplier \cite{usman2023low}.}
 	\label{fig: osppm}
\end{figure}

The design of the LR serial-parallel multiplier (LR-SPM) has been presented in \cite{usman2023low} and is depicted in Fig.~\ref{fig: osppm}. The algorithm of the multiplier contains an \textit{initialization stage}: having a execution length equal to $\delta$ during which the input digits are collected and no output is generated and a \textit{recurrence stage}: which executes for $j$ iterations, producing one output digit in each iteration. The multiplication algorithm is clearly laid out in Algorithm~\ref{alg:alg1}. The LR-SPM has an online delay $\delta$ of $2$. The derivation of LR-SPM can be found in \cite{usman2023low}.

\begin{algorithm}[H]
\caption{Radix-2 LR serial-parallel multiplication algorithm}\label{alg:alg1}
\begin{algorithmic}
\STATE $\textbf{procedure }\; \text{LR-SPM} (X, Y, P_{\text{out}})$
\STATE $X$: Parallel input (Element of input feature map)
\STATE $Y$: Serial input (Element of filter)
\STATE $P_{\text{out}}$: Serial output
\STATE 1. [Initialize]
\STATE \hspace{0.5cm} $y[-2] = w[-2] = 0$
\STATE \hspace{0.5cm}$\textbf{for } (j = -2,-1) \textbf{ do}$
\STATE \hspace{1cm}$v[j] = 2w[j] + (X.y_{j+2})2^{-2}$
\STATE \hspace{1cm}$w[j+1] \leftarrow v[j]$
\STATE \hspace{0.5cm}$\textbf{end for}$
\STATE 2. [Recurrence]
\STATE \hspace{0.5cm}$\textbf{for } (j = 0,....,n-1) \textbf{ do}$
\STATE \hspace{1cm}$v[j] = 2w[j] + (X.y_{j+2})2^{-2}$
\STATE \hspace{1cm}$p_{j+1} = \text{SELM}(\widehat{V[j]})$
\STATE \hspace{1cm}$w[j+1] \leftarrow v[j] - p_{j+1}$
\STATE \hspace{1cm}$P_{\text{out}} \leftarrow p_{j+1}$
\STATE \hspace{0.5cm}$\textbf{end for}$
\STATE $\textbf{end procedure}$
\end{algorithmic}
\end{algorithm}

\begin{figure}[ht]
	\begin{center}
 		\includegraphics*[width=4cm]{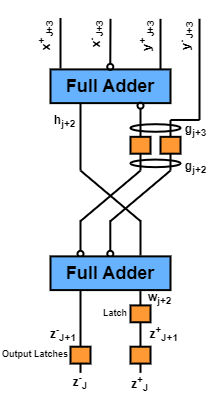}
 	\end{center}
 	\caption{Radix-2 LR Adder \cite{ercegovac2004digital}.}
 	\label{fig: OA}
 \end{figure}

In computation modules where LR arithmetic-based multipliers are used, the generation of the sum of products (SoP) requires compatible adder modules, such as LR adders. The LR arithmetic-based adders offer a unique advantage of precision independent addition over conventional adders such as carry-propagate adder, carry save adder, etc. 
An MSDF digit-serial LR adder is adopted in this work to receive inputs and produce outputs in a MSDF fashion. The LR adder used in this study has an online delay of \(\delta = 2\), and is illustrated in Fig.~\ref{fig: OA}. Further details and derivations of the LR adder can be found in \cite{ercegovac2004digital}.

\section{Proposed Design: DSLR-CNN Design} \label{sec: DSLR-CNN}

In this section, we present a comprehensive overview of the proposed design, which leverages LR arithmetic-based SoP units. The proposed design is composed of computation units, input and activation buffers, and a control unit (CU), are carefully designed and organized to optimize performance. Convolution operations involve MAC operations, entailing the multiplication and addition of input values with weight values. To streamline the convolution process, the proposed design incorporates tiling factors as mentioned in Table~\ref{tab:conf} . These elements are specifically tailored to accommodate $9$ processing elements (PE) in each column and $64$ PEs in a row, each equipped with an input tiling factor of $16$, and output tiling factor of $8$ to produce the output feature map. As a result, a total of $16 \times 9 \times 64 \times 8$ PE units facilitates convolution operations. 

\subsection{Overall Design}
\begin{figure*}[ht]
	\begin{center}
 		\includegraphics*[width=16cm]{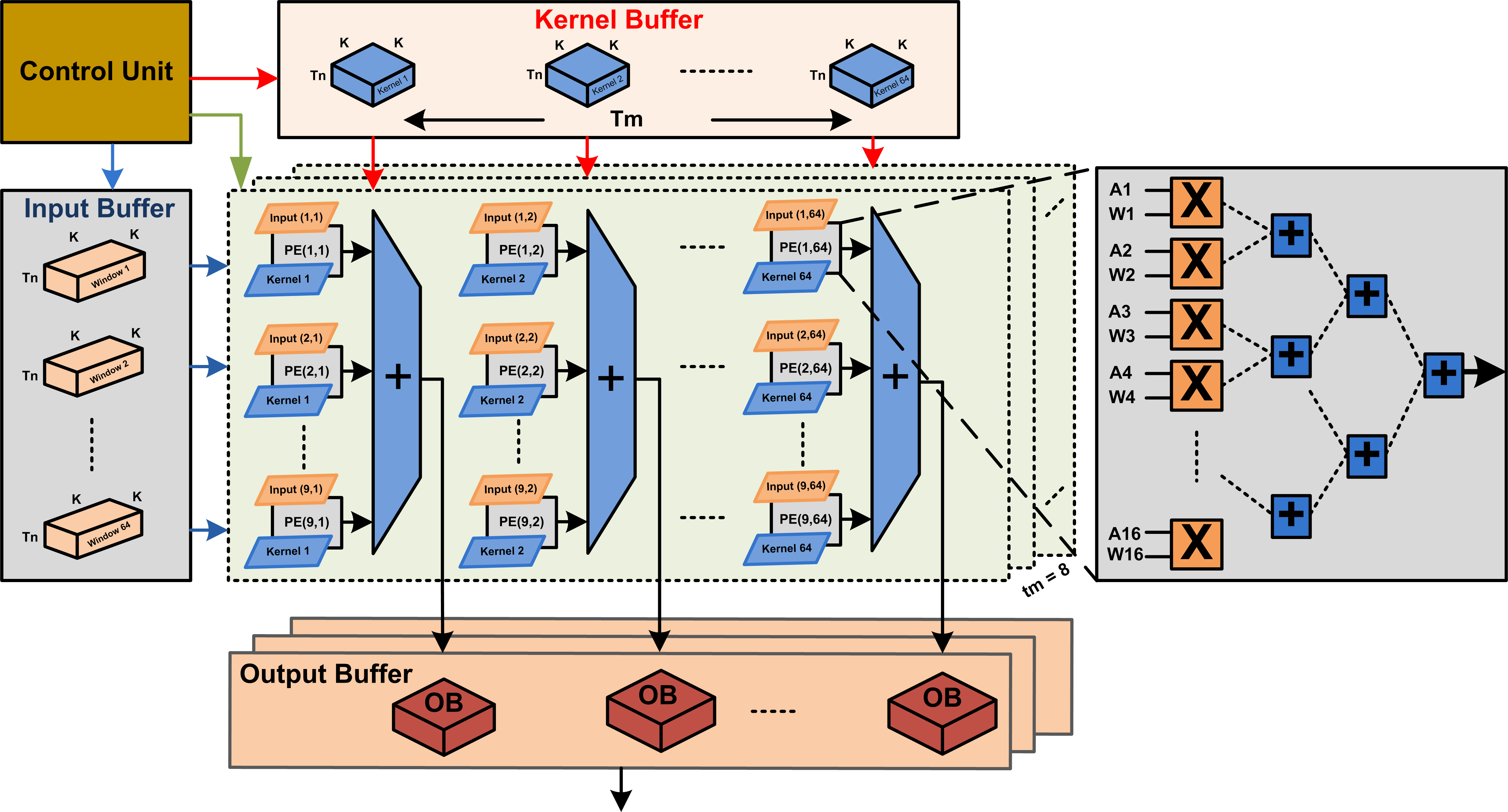}
 	\end{center}
 	\caption{Tile of the DSLR-CNN Architecture and its Processing Element.}
 	\label{fig: OD}
 \end{figure*}

 The proposed DSLR-CNN design comprises of several essential components such as PEs, control unit, and input activation/filter buffers as depicted in Fig.~\ref{fig: OD}. The PEs are the primary units that perform convolution computations.

In the DSLR-CNN architecture, we employ a tiling technique, a form of data processing parallelism designed to maximize hardware resources. Specifically, we introduced two key tiling factors: input tiling denoted as $(T_{n})$, set to $16$, and output tiling $(T_{m})$ set to $8$ aimed at improving the efficiency of convolution operations. Here, $T_{m}$ represents the number of tiling factors for output channels while $T_{n}$ indicates the number of tiling factors for input channels of a convolution layer. Additionally, we employ tiling factors for rows and columns, represented by $T_{r}$ and $T_{c}$, each set to $8$, totaling $64$, to optimize resource utilization by breaking down input and output data into manageable chunks. Initially, the inputs and kernels of a layer from the pre-trained model, each with a size of $k \times k$, are stored in off-chip memory. Here, the weight of the layer is written onto the kernel buffer, with each bank designated to hold one weight filter.  All the PEs in a column of a DSLR-CNN tile ensure a similar distribution of kernels. Similarly, the rows of PEs are connected to the input buffer, where each bank accommodates one window of the input feature map. This design guarantees that all PEs within the same row receive the same input feature map. The control unit manages this data flow and configures the accelerator accordingly. The architecture comprises of rows, columns, input, and output channel tiling. With $9$ PEs per row and $64$ PEs per column, input channel tiling of $16$, and output channel tiling of $8$ , facilitating the online processing of $9 \times 64 \times 16 \times 8$ convolution windows in each cycle. Within each PE, input activation is serially processed one bit at a time, while the kernel data is fed in parallel. Each PE generates partial output by convolving a weight filter with an input window, with the LR-SPM unit computing the convolution operation by multiplying corresponding pixels. Subsequently, a reduction tree aggregates these convolution products channel-wise, resulting in final output pixels written to the output buffer. This iterative process continues until all $9 \times 64 \times 16 \times 8$ kernels have been convolved with the input feature map, enabling parallel processing across all input channels. The number of cycles required for the proposed design to produce its output is determined using a formulated approach defined by \eqref{eq3}, ensuring efficient and timely computation throughout the network.

\begin{equation}\label{eq3}
\begin{split}
N_{\text{Cycles}} = \biggl(\delta_{\text{mult}} + \delta_{\text{add}} \times \Big\lceil \log_{2} (k \times k) \Big\rceil + \delta_{add} \times \Big\lceil \log_{2}(T_{n}) \Big\rceil 
\\+ P_{i} +  \Big\lceil \log_{2}(k \times k) \Big\rceil + \Big\lceil \log_{2}(T_{n}) \Big\rceil \biggl)\\ \times  \Big\lceil \frac{R \times C}{T_{r} \times T_{c}}\Big\rceil  \times \Big\lceil \frac{M}{T_{m}} \Big\rceil\times \Big\lceil \frac{N}{T_{n}}\Big\rceil
\end{split}
\end{equation}

$\delta_{add}$ and $\delta_{mult}$ represent the online delays of the online adder and multiplier respectively. $\lceil \log_{2} (k \times k)\rceil$ signifies the number of reduction stages in the adder tree required to generate the SoP of the $k \times k$ multipliers. $T_{n}$ denotes the input tiling, while $P_{i}$ indicates the input precision. The depths of the channel reduction tree are given by $\lceil \log_{2} (k \times k) \rceil$ and $\lceil \log_{2}(T_{n}) \rceil$. Furthermore, $ \lceil \frac{R \times C}{T_{r} \times T_{c}}\rceil$ represents the tiling of rows and columns. $\lceil \frac{M}{T_{m}} \rceil$ referred to as number of output feature map and the output tiling, in the proposed design, while $\lceil \frac{N}{T_{n}}\rceil$ denotes the input feature map and input tiling for the DSLR-CNN architecture.

 \subsection{Processing Element}\label{PE}
  The architecture of PE design as shown in Fig.~\ref{fig: OD}, comprises of $16$ LR serial-parallel multipliers followed by an online adder tree to perform the convolution operation for a given input channel.  In the proposed design, we employ a weight stationary dataflow approach within the PEs. This employs that the weights (or kernel values) remain fixed within the PE during the computation of the entire output feature map, while the input pixels are streamed serially in a MSDF fashion. The kernel pixel is fed in parallel, allowing for efficient reuse of the weights, which minimizes data movement and enhances computation efficiency. Each multiplier in the PE is responsible for the multiplication of one pixel in the convolution window with the corresponding pixel in the same feature map of the convolution kernel. Therefore, $16$ pixels can be processed in parallel, in each PE. For instance, when an $n$-bit activation and weight perform a convolution operation, the MSB of the product is generated after $\delta=2$ cycles, and the complete output is generated in $n + \delta$ cycles, where $n$ is the precision of the output. The convolution output is generated by feeding the digit-serial product from the multipliers directly to the online reduction tree. The reduction tree then performs the sum of these $k\times k$ products and generates the result of the convolution in an MSDF manner.

\subsection{Control Unit}
\begin{figure}[ht]
	\begin{center}
 		\includegraphics*[width= 7cm]{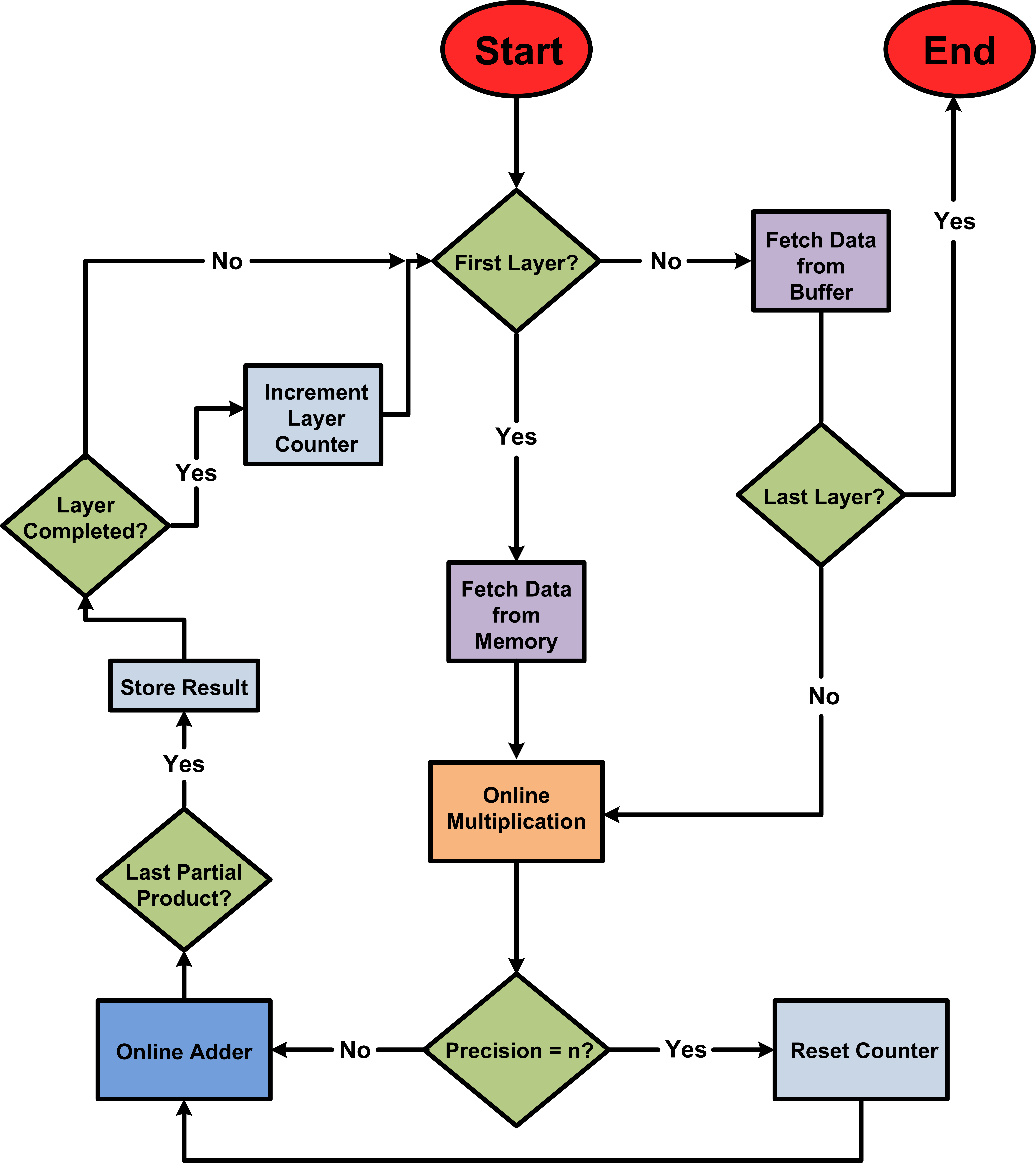}
 	\end{center}
 	\caption{Flowchart of the Control Unit.}
 	\label{fig: flow}
 \end{figure}

The control unit in Fig.~\ref{fig: OD} generates control signals for the different components based on the requirements of the current layers. The control unit holds essential information about the CNNs architecture, such as the number of layers, filters, and the size of the input and output feature maps. The CU utilizes this information to generate the control signals for configuring the accelerator. Additionally, it constantly monitors the accelerators execution. If the accelerator encounters an error, the CU generates a new control signal to remedy it. The CU is an integral component of the DSLR-CNN architecture, ensuring that the CNN accelerator is accurately configured and executes the workload normally. The CU is also responsible for managing event triggers to facilitate smooth data flow among various modules based on system states. This involves coordinating start/end signals, updating variables, managing read/write operations, and selecting data sources. As shown in Fig.~\ref{fig: flow}, this process begins with the CU initiating the inference process by fetching data from external memory. Once the data is fetched, it is sent to PEs in a serial manner, indicating that the data is processed bit by bit, in LR fashion. After the data reaches the PEs, the online multiplication operation is triggered. This multiplication is crucial for calculating the outputs of each neural network layer. The control unit checks whether the required precision $n$ for this multiplication is achieved. If not, the system continues the multiplication process and when the desired precision is achieved, the counter is reset to prepare for the next set of operations, ensuring that no residual data interferes with future calculations. Next, the control unit directs the system, to proceed to the online addition stage if the last partial product has not been reached. This step is important for combining the partial results generated by the online multiplication process. Once the addition is complete, the control unit  stores the results and check whether the computation of the current layer is completed. If it is finished, the CU determines whether it is the last layer of the network. If the last layer has not been reached, the control unit increments the layer counter and repeats the process for the next layer. This loop continues until all layers are processed. Finally, when the last layer is completed, the process ends.

\subsection{ Input and Kernel Buffer}

The control unit carefully manages the data flow between on-chip memory and the input/kernel buffers, ensuring prompt access when needed. The input/kernel buffer interfaces with the CU, which holds the input activations and kernels (weights), for the executing layers. Each PE simultaneously handles an identical window of the input feature map, enabling parallel processing. Data exchange between buffers and PEs occurs through interconnected wires. The CU oversees communication between PEs and buffers, delivering input activation in bit-serial format and constant-weight values to PE rows. It carefully monitors the accurate storage of intermediate computation results in the output buffer. Typically located on-chip, this buffer offers high bandwidth, allowing other accelerator components to easily access and store intermediate computations when needed.

\section{Results and Discussion} \label{sec: Results}

This section provides a comprehensive overview of the experimental setup, performance evaluation metrics, comparison, and result of the DSLR-CNN design.

\subsection{Performance Evaluation}
The performance of hardware accelerators for DNN acceleration is often influenced by the specific target application. However, researchers have established several standard metrics to evaluate these accelerators \cite{ chen2016eyeriss,cheng2024leveraging,hsu2020essa, samimi2019res, lu2021distilling}. Key metrics include the number of computation cycles, power consumption, area, throughput, latency, power efficiency, and overall performance. These metrics are essential to compare and assess the strengths and limitations of different hardware designs, helping to develop and deploy DNNs.

\subsubsection{Power Utilization and Energy Efficiency}

Power utilization refers to the amount of energy consumed over a specific period. Power consumption is typically reported in milliwatts (mW) or joules per second. Energy efficiency, on the other hand, measures how much data can be processed or the number of tasks can be completed per unit of energy. This is especially crucial when running DNNs on embedded devices at the edge. Energy efficiency is typically expressed as the number of operations performed per joule. For inference tasks, energy efficiency is often measured as GOPS per watt (GOPS/W) or TOPS per watt (TOPS/W), while energy consumption is quantified as joules per inference.

\subsubsection{Area and Area Efficiency}
The area refers to the amount of silicon required for the DNN acceleration, typically measured in square millimeters or micrometer square ($\mu m^2$) . The area efficiency can be quantified by assessing system performance in GOPS per millimeter square (GOPS/$mm^{2}$) or TOPS per millimeter square (TOPS/$mm^{2}$). The area required for DNN acceleration is influenced by on-chip memory size and the technology used in hardware synthesis. The total available area and the size of each PE are crucial in determining the number of PEs that can be integrated into the system. To increase the number of PEs, without expanding the area, two main strategies are employed: reducing PE size and on-chip storage. Reducing PE size involves minimizing the space used by its logic or components, allowing more PEs to fit within the same area. Alternatively, reducing on-chip memory allows for additional PEs, but this may negatively affect PE utilization efficiency, as lower memory can slow data access and impact performance. Additionally, the per-PE area can be minimized by simplifying the logic for data transmission to the MAC units, which enables integrating more PEs within the same area but may involve performance trade-offs \cite{sze2020evaluate}.


\subsubsection{Throughput or Performance}
Throughput is the rate at which data is processed or tasks are completed in a given time frame. It is a crucial metric for evaluating the efficiency of network connections or data processing systems. Typically measured in giga operations per second (GOPS), or in tera operations per second (TOPS), higher throughput suggests a more efficient network or system \cite{wang2019deep}. The performance of the proposed design can be assessed using \eqref{performance},

\begin{equation}\label{performance}
    {Performance} = \frac{{Number \;of\;operations}} {{Duration (ms)}}
\end{equation}

The equation to determine the total number of operations (OPS) for a given convolution layer is derived by the formula $2 \times M \times N \times R \times C \times K \times K$. Where, $M$ and $N$ represent the number of output and input feature maps, while $R$ and $C$ represent the height and width of the output feature map. The term $K \times K$ signifies the dimension of the convolution kernel.

\subsubsection{Latency}
Latency or duration measures the time taken from the arrival of input data at a system to the time the result is generated. In a network, latency and throughput can be derived from each other. Applications that rely on real-time interaction such as augmented reality, autonomous navigation, and robotics require low latency to function effectively. However, as latency increases, it can limit the maximum achievable throughput in a data exchange between two points. Because of this relationship, achieving both high throughput and low latency can be challenging and sometimes mutually exclusive, making it important to report both metrics \cite{wang2019deep}. Latency is typically measured in milliseconds (ms), or in nanoseconds (ns). The equation to compute the duration of the proposed design is given in \eqref{duration},

\begin{equation}\label{duration}
    Duration  = \frac{Execution_{Cycles}}{Frequency (Mhz)}
\end{equation}

\subsubsection{Analysis}

\begin{figure}[ht]
	\begin{center}
 		\includegraphics*[width= 8cm]{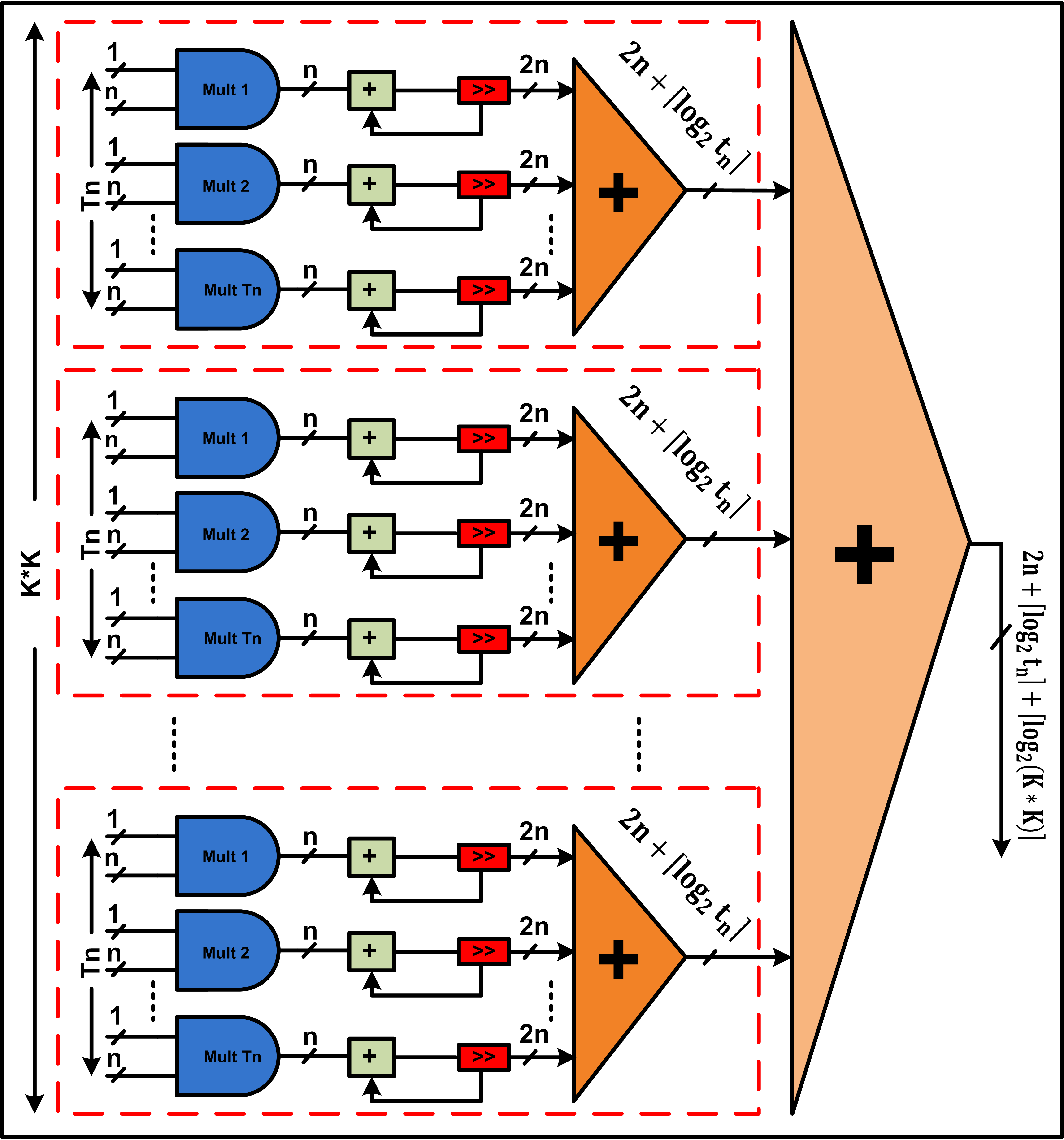}
 	\end{center}
 	\caption{Baseline Design: Conventional Bit-Serial Architecture Used for Comparative Analysis.}
 	\label{fig: unpu}
 \end{figure}

 The analysis of the hardware accelerator begins with a comparison of synthesis results between the proposed design and a conventional bit-serial baseline, both of which were developed and implemented by us in RTL (Verilog) as shown in Fig.~\ref{fig: unpu}. The baseline follows the same dataflow architecture and array layout as the proposed design, ensuring a consistent comparison using 16-bit input precision. The architecture of the conventional bit-serial design contains AND gate arrays for partial product generation, followed by an accumulator to obtain the sum of partial products as shown in Fig.~\ref{fig: unpu}. The accumulation process presented in the figure is followed by an adder tree to perform the sum of $k \times k$ products. The RTL designs were synthesized using the GSCL 45nm cell library at a nominal supply voltage of $1.1$ V and evaluated at a clock frequency of $500$ MHz. The comparison of synthesis focuses on key performance metrics such as latency, area utilization, and power consumption , as presented in Table~\ref{tab:syn}, along with the critical path delay for each design. To further evaluate and compare the performance of the DSLR-CNN with the baseline design, we conducted a comprehensive evaluation of the proposed design across three networks AlexNet \cite{krizhevsky2012imagenet}, VGG-16 \cite{simonyan2014very}, and ResNet-18 \cite{he2016deep} using the ImageNet dataset \cite{deng2009imagenet}. Notably, AlexNet, VGG-16, and ResNet-18 feature $5$, $13$, and $17$ convolution layers, respectively. 
This evaluation measures total duration, peak performance, peak energy efficiency, and peak area efficiency, quantified in ms, TOPS, TOPS/W, and GOPS/$mm^2$ respectively, as shown in Table~\ref{tab:alex}. The results demonstrate that DSLR-CNN outperforms the baseline, demonstrating its robustness and efficiency in accelerating CNN computations. The execution cycle computation for the baseline design is provided in equations \eqref{eq5}. Additionally, Fig.\ref{fig: roof} provides an analysis of performance (TOPS) and operational intensity (TOPS/Byte). The proposed design is further compared to previous works as depicted in Table~\ref{tab:compare}.


\begin{equation}\label{eq5}
\begin{split}
    Cycle_{Base} =  \Big(( Mult + Acc) \times n + \lceil \log_2(T_{n}) \rceil \\
    + \lceil \log_2(K^{2}) \rceil \Big)  \times \Big \lceil \frac{R}{T_{r}} \Big \rceil \times \Big \lceil \frac{C}{T_{c}} \Big \rceil\times \Big \lceil \frac{M}{T_{m}} \Big \rceil \times \Big \lceil \frac{N}{T_{n}} \Big \rceil
     \end{split}
\end{equation}

where, $ Mult +Acc+ \lceil\log_{2}(T_{n})\rceil  $ denotes the $AND$ multiplier array, accumulator, and tiling factor of the number of input channels to perform the MAC operation, $n$ is the precision, $K^{2}$ are the convolution kernels sizes. The factors $R$, $C$, and $M$ are the rows, columns, and output filters, respectively. Furthermore, $T_{m}$, $T_{r}$, and $T_{c}$ are the factors of output, row, and column tiling.

\subsubsection{Synthesis Results Comparison}
We compared and analyzed the synthesis results of the DSLR-CNN design with the baseline, utilizing GSCL 45nm technology, at a frequency of $500$ MHz. The comprehensive findings are summarized in Table~\ref{tab:syn}. The DSLR-CNN design, employing 16-bit precision level showcased remarkable performance in terms of latency, surpassing the baseline that employs PE array design. Specifically, DSLR-CNN achieved a latency of $1.07$, marking a substantial improvement over the baseline design. This reduction in latency is attributed to the dependency of the proposed design on input precision and online delay. Particularly, in an LR algorithm, the inter-operation latency is solely dependent on $\delta$ only, which remain fixed and small. As multiple operations are performed online, the total delay is the sum of the delays for each operation. Notably, the precision of the calculation does not affect this delay \cite{usman2023low} making LR arithmetic algorithms highly effective for wider word sizes and extensive data-dependent operations. Furthermore, it is imperative to acknowledge that the proposed design exhibited high power consumption and area utilization compared to the baseline. This outcome is attributed to the intricacies of our dataflow design, which prioritizes performance optimization through robust parallelism; Featuring $9$ PEs per column and $64$ PEs per row, each equipped with $16$ multipliers. However, this optimization strategy comes with the trade-off of increased power consumption and larger area utilization. Despite these challenges, our dataflow-centric architecture remains pivotal in achieving superior performance metrics for CNN accelerators. Furthermore, we analyzed the critical path of the proposed design as the sum of the critical path of the LR multiplier and the subsequent reduction tree as shown in \eqref{eq6} and \eqref{eq7}. For baseline, the critical path delay is provided in \eqref{eq8}.

\begin{table}[ht]
\centering
\renewcommand{\arraystretch}{1}
\caption{Synthesis Results of the DSLR-CNN Accelerator Compared to the Baseline Using GSCL 45nm Technology.}\label{tab:syn}
\resizebox{0.8\linewidth}{!}{
\begin{tabular}{|l|c|c|c} \hline 
\textbf{Parameter} & \textbf{Baseline}   & \textbf{DSLR-CNN
} \\ \hline \hline

Latency (ns) 
& 1.92
& 1.07  \\ \hline

Area ($\mu m^2$) 
& 54,206,087
& 84,046,898    \\ \hline

Power (mW) 
& 795.21
& 1249.42 \\ \hline  
\end{tabular}
}
\end{table}

 \begin{equation}\label{eq6}
     t_{OLM}= t_{MUX{2:1}} + t_{Adder{3:2}}+t_{CPA-4}+t_{SELM}+t_{XOR}
 \end{equation}

The critical path of an online adder (OLA) is found to be
\begin{equation}\label{eq7}
    t_{OLA}=2 \times t_{FA}+t_{FF}
\end{equation}

The critical path for the baseline shown in relation \eqref{eq8}, includes the delays of $AND$ gate, followed by an accumulator ($ADD-16$) and carry propagate adder ($CPA-32$) to add the output of $16$ $AND$ gates, and finally another CPA referred to as, $t_{CPA-36}$, to perform the reduction of the results obtained from the reduction tree referred as $CPA-32$.   
\begin{equation}\label{eq8}
t_{baseline} = t_{AND} +t_{ADD-16}+t_{CPA-32}+t_{CPA-36}
\end{equation}

\subsubsection{Performance Comparison}

In our initial experiment, we conducted a comprehensive evaluation of the proposed DSLR-CNN across three networks: AlexNet, VGG-16, and ResNet-18. The layer-wise architecture of these CNN models is presented in Table~\ref{tab:models}. Operating at a frequency of 500 MHz, we thoroughly assessed various design performances on these networks, focusing on total duration, peak performance, peak energy efficiency, and peak area efficiency, as detailed in Table~\ref{tab:alex}. The performance comparison between DSLR-CNN and the baseline reveals insightful trends. For AlexNet, the DSLR-CNN design significantly outperforms the baseline configuration, achieving a peak performance of $4.47$ TOPS, while maintaining a lower total duration of $0.94$ ms, across the convolution layers. Moreover, it exhibits an improved energy efficiency of $3.57$ TOPS/W and a significantly increased area efficiency at $53.18$ GOPS/$mm^2$. In the case of VGG-16, the DSLR-CNN consistently shows higher performance, with a peak performance of $1.75$ TOPS and a significantly lower total duration of $1.44$ ms. This improvement is coupled with an energy efficiency of $1.40$ TOPS/W and an area efficiency of $20.82$ GOPS/$mm^2$, further highlighting the proposed design superior efficiency. For ResNet-18, the DSLR-CNN demonstrates significant performance advantages, achieving a peak performance of $1.75$ TOPS and a remarkably lower total duration of $0.13$ ms. The energy efficiency is improved at $1.40$ TOPS/W, and the area efficiency stands at $20.82$ GOPS/$mm^2$, consistently outperforming the baseline across all metrics.

\renewcommand{\arraystretch}{1.4}
\begin{table}[!ht]
    \centering
    \caption{Convolution Layer Architecture of AlexNet, VGG-16, and ResNet-18 Networks,  Where \(M\) Represents the Number of Kernels (Output Feature Maps) and \(R\times C\) Denotes the Dimensions of the Output Feature Maps.}
    \begin{tabular}{l l c c c} \hline
        \textbf{Network} & \textbf{Layer} & \textbf{Kernel Size} & \(M\)  & \(R \times C\) \\ \hline \hline
         \multirow{5}{*}{AlexNet} & C1 & {\(11 \times 11\)} & 96  & \(55 \times 55\) \\
        & C2 & \multirow{1}{*}{\(5 \times 5\)} & 256  & \(27 \times 27\) \\
        & C3 & \multirow{1}{*}{\(3 \times 3\)} & 384  & \(13 \times 13\) \\
        & C4 & & 384  & \(13 \times 13\) \\
        & C5 & & 256  & \(13 \times 13\) \\ \hline
        \multirow{5}{*}{VGG-16} & C1-C2 & \multirow{5}{*}{\(3 \times 3\)} & 64  & \(224 \times 224\) \\
        & C3-C4 & & 128  & \(112 \times 112\) \\
        & C5-C7 & & 256  & \(56 \times 56\) \\
        & C8-C10 & & 512  & \(28 \times 28\) \\
        & C11-C13 & & 512  & \(14 \times 14\) \\ \hline
        \multirow{5}{*}{ResNet-18} & C1 & \(7\times 7\) & 64 & \(112 \times 112\) \\
        & C2-C5 & \multirow{4}{*}{\(3 \times 3\)} & 64 & \(56 \times 56\) \\
        & C6-C9 & & 128 & \(28 \times 28\) \\
        & C10-C13 & & 256 & \(14 \times 14\) \\
        & C14-C17 & & 512 & \(7 \times 7\) \\ \hline
    \end{tabular}
    
    \label{tab:models}
\end{table}

\begin{table*}[ht!]
\centering
\caption{Performance Comparison Between the DSLR-CNN Design and the Baseline on AlexNet, VGG-16, and ResNet-18 Networks. Both Designs Employ Similar Dataflow and Configuration, Implemented in 45nm Technology at 500 MHz. Metrics Include Total Duration i.e., total inference time/layer (ms), Peak Performance (TOPS), Peak Energy Efficiency (TOPS/W), and Peak Area Efficiency (GOPS/$mm^{2}$). } \label{tab:alex}
\resizebox{0.8\textwidth}{!}{
\renewcommand{\arraystretch}{1.4}
\begin{tabular}{|cc|c|c|c|}
\hline
\multicolumn{2}{|c|}{\textbf{Network}}                                                    & \textbf{AlexNet} & \textbf{VGG-16} & \textbf{RESNET-18} \\ \hline
\multicolumn{2}{|c|}{\textbf{Layers}}                                                     & Conv 1-5         & Conv 1-13       & Conv 1-17          \\ \hline
\multicolumn{1}{|c|}{\multirow{4}{*}{\textbf{Baseline}}} & \text{Total Duration (ms)}   & 1.54             & 2.40             & 0.23               \\ \cline{2-5} 
\multicolumn{1}{|c|}{}                                   & \text{Peak Perf. (TOPS)}     & 2.73             & 1.05            & 1.05               \\ \cline{2-5} 
\multicolumn{1}{|c|}{}                                   & \text{Peak Energy Eff. (TOPS/W)}   & 3.43             & 1.32            & 1.32               \\ \cline{2-5} 
\multicolumn{1}{|c|}{}                                   & \text{Peak Area Eff. (GOPS/$mm^2$)} & 50.39            & 19.37           & 19.37              \\ \hline
\multicolumn{1}{|c|}{\multirow{4}{*}{\textbf{DSLR-CNN}}} & \text{Total Duration (ms)}   & 0.94             & 1.44            & 0.13               \\ \cline{2-5} 
\multicolumn{1}{|c|}{}                                   & \text{Peak Perf. (TOPS)}     & 4.47             & 1.75            & 1.75               \\ \cline{2-5} 
\multicolumn{1}{|c|}{}                                   & \text{Peak Energy Eff. (TOPS/W)}   & 3.57             & 1.40             & 1.40                \\ \cline{2-5} 
\multicolumn{1}{|c|}{}                                   & \text{Peak Area Eff. (GOPS/$mm^2$)} & 53.18            & 20.82            & 20.82               \\ \hline
\end{tabular}}
\end{table*}

\begin{figure}[ht]
    \centering
    \begin{subfigure}[h]{1\linewidth}
        \centering
        \includegraphics[width=\linewidth]{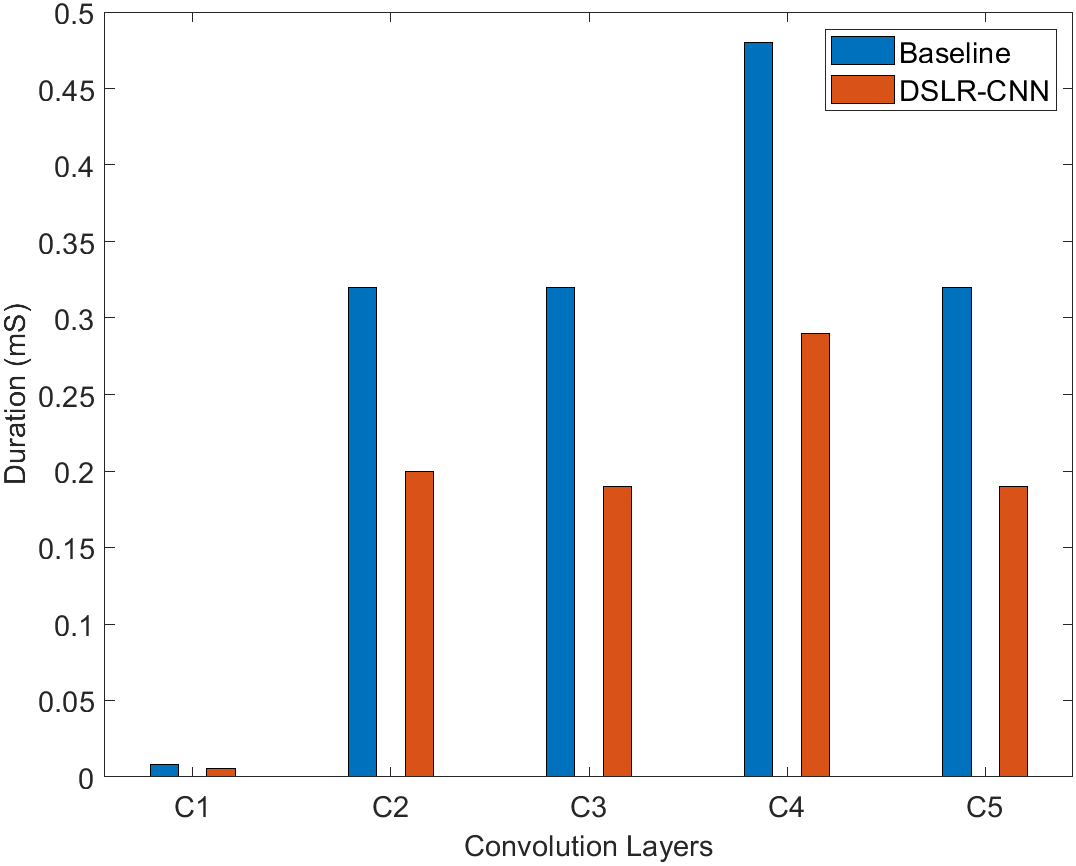}
        \caption{   }
    
    \end{subfigure}
    \vspace{0.5cm}  
    \begin{subfigure}[h]{1\linewidth}
        \centering
        \includegraphics[width=\linewidth]{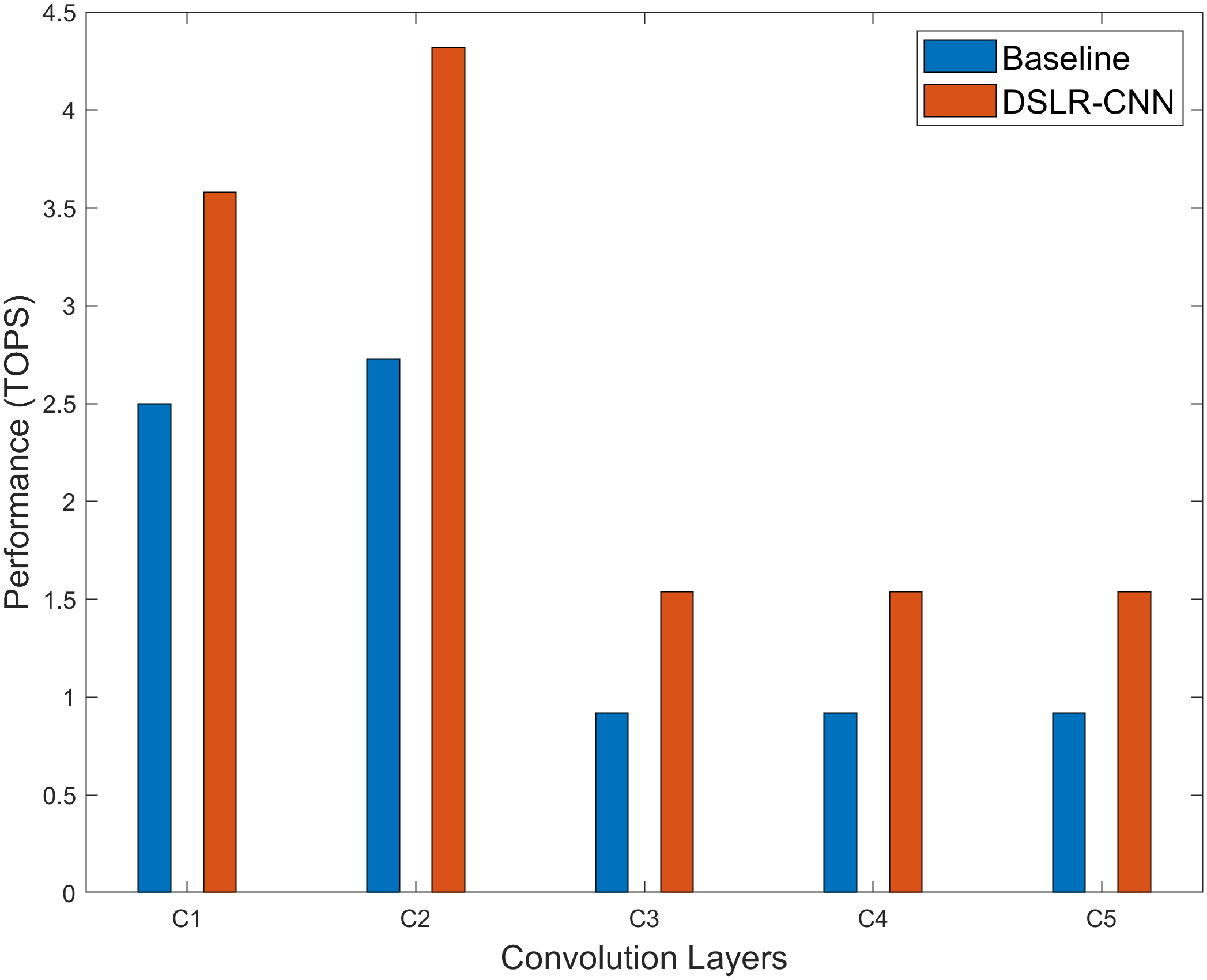}
        \caption{   }
        
    \end{subfigure}

    \caption{Comparison of AlexNet Network: (a) Duration and (b) Performance.}
     \label{fig:alexcomparison}
\end{figure}

These findings underscore the potential of the proposed digit-serial LR arithmetic-based design to accelerate and optimize processing in complex neural network architectures. The substantial enhancements in performance and inference time, as compared to the baseline design, illustrate the efficacy of the proposed DSLR-CNN design for deep learning applications. Fig.~\ref{fig:alexcomparison} illustrates the performance and duration of the convolution layers within the AlexNet network, showcasing substantial enhancements in performance and inference time compared to the baseline design.

\begin{figure}[ht]
    \centering
    \begin{subfigure}[h]{1\linewidth}
        \centering
        \includegraphics[width=\linewidth]{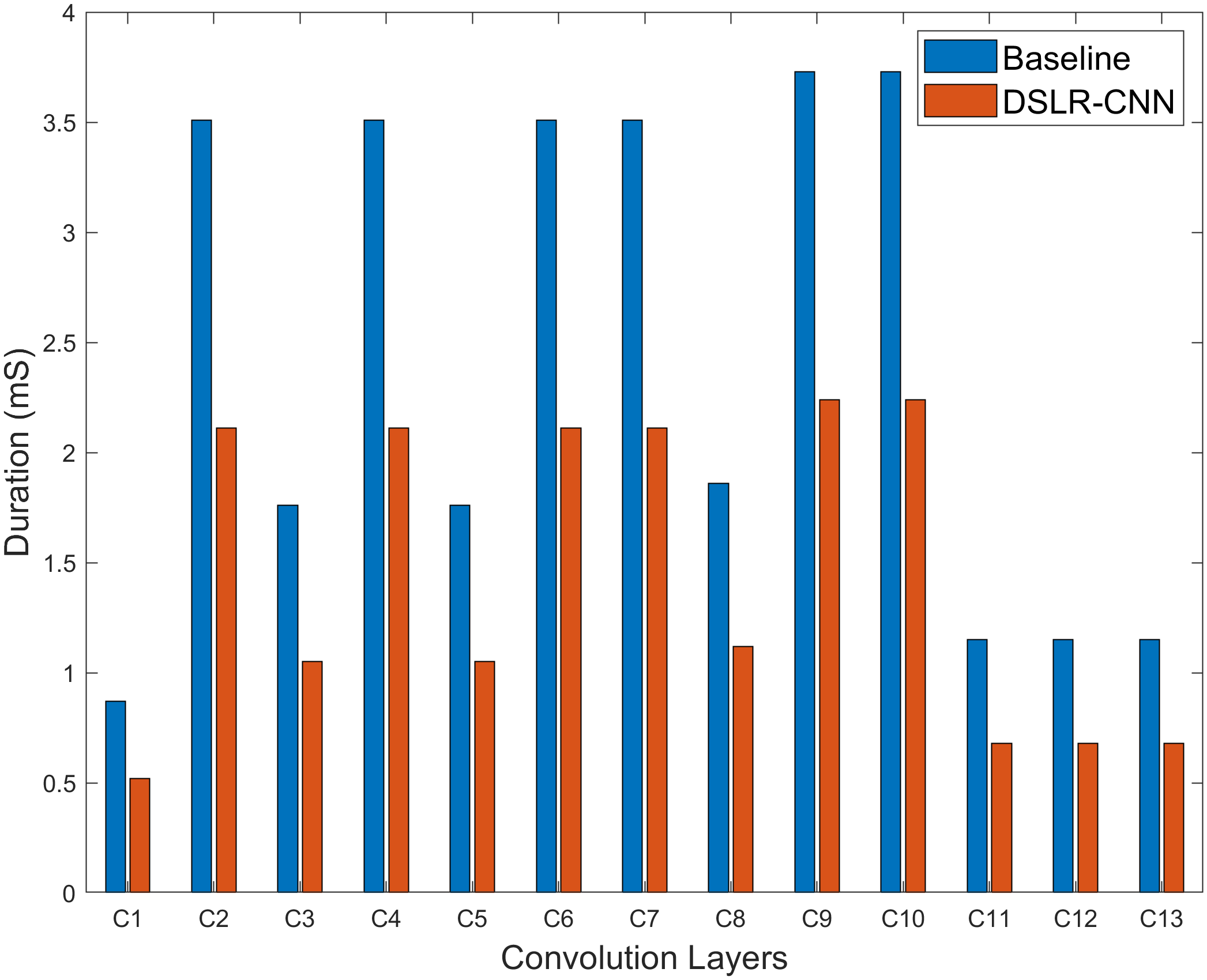}
        \caption{   }
    
    \end{subfigure}
    \vspace{0.5cm}  
    \begin{subfigure}[h]{1\linewidth}
        \centering
        \includegraphics[width=\linewidth]{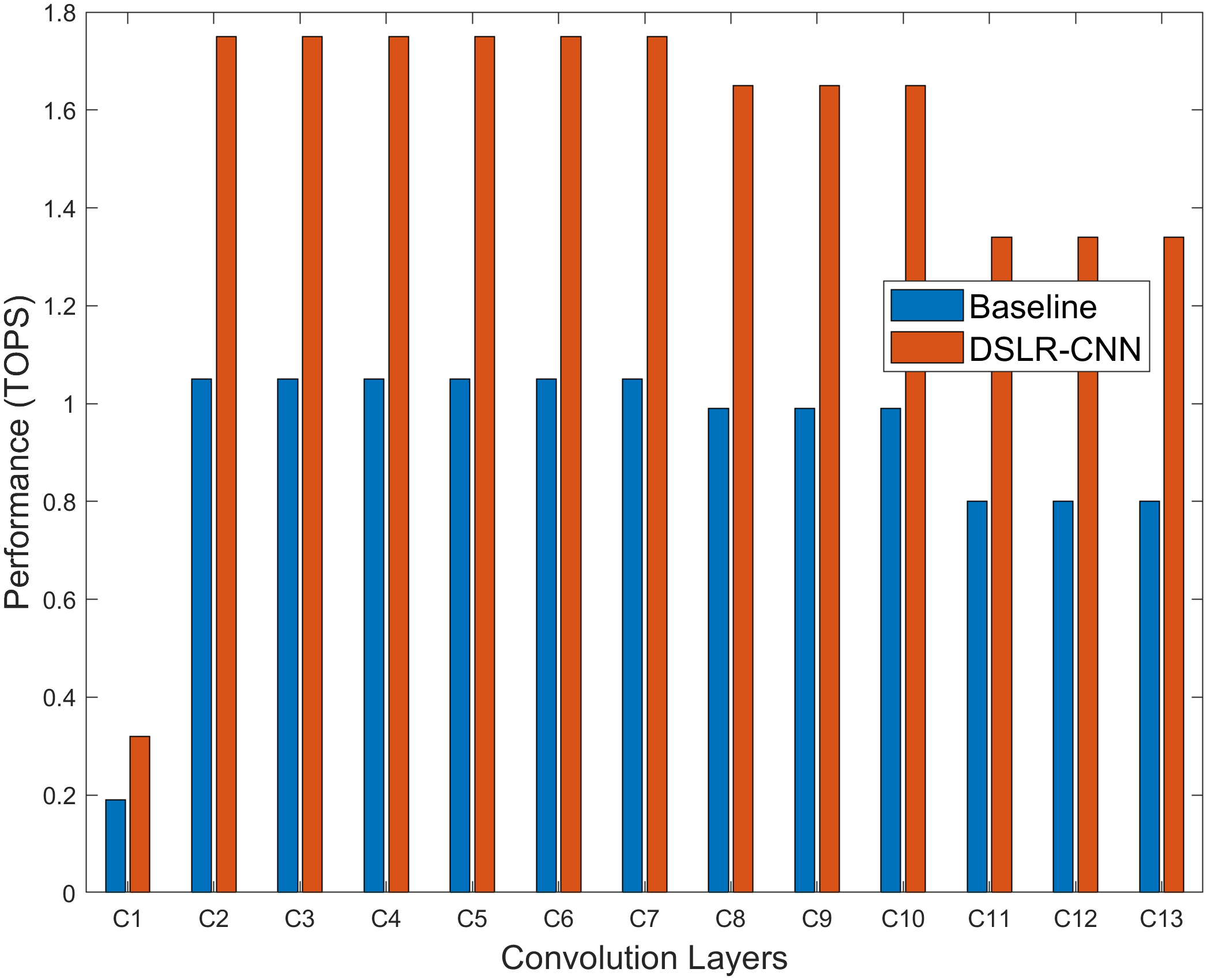}
        \caption{   }
        
    \end{subfigure}

    \caption{Comparison of VGG-16 Network: (a) Duration and (b) Performance.}
     \label{fig:vggcomparison}
\end{figure}

\begin{figure}[ht]
    \centering
    \begin{subfigure}[h]{1\linewidth}
        \centering
        \includegraphics[width=\linewidth]{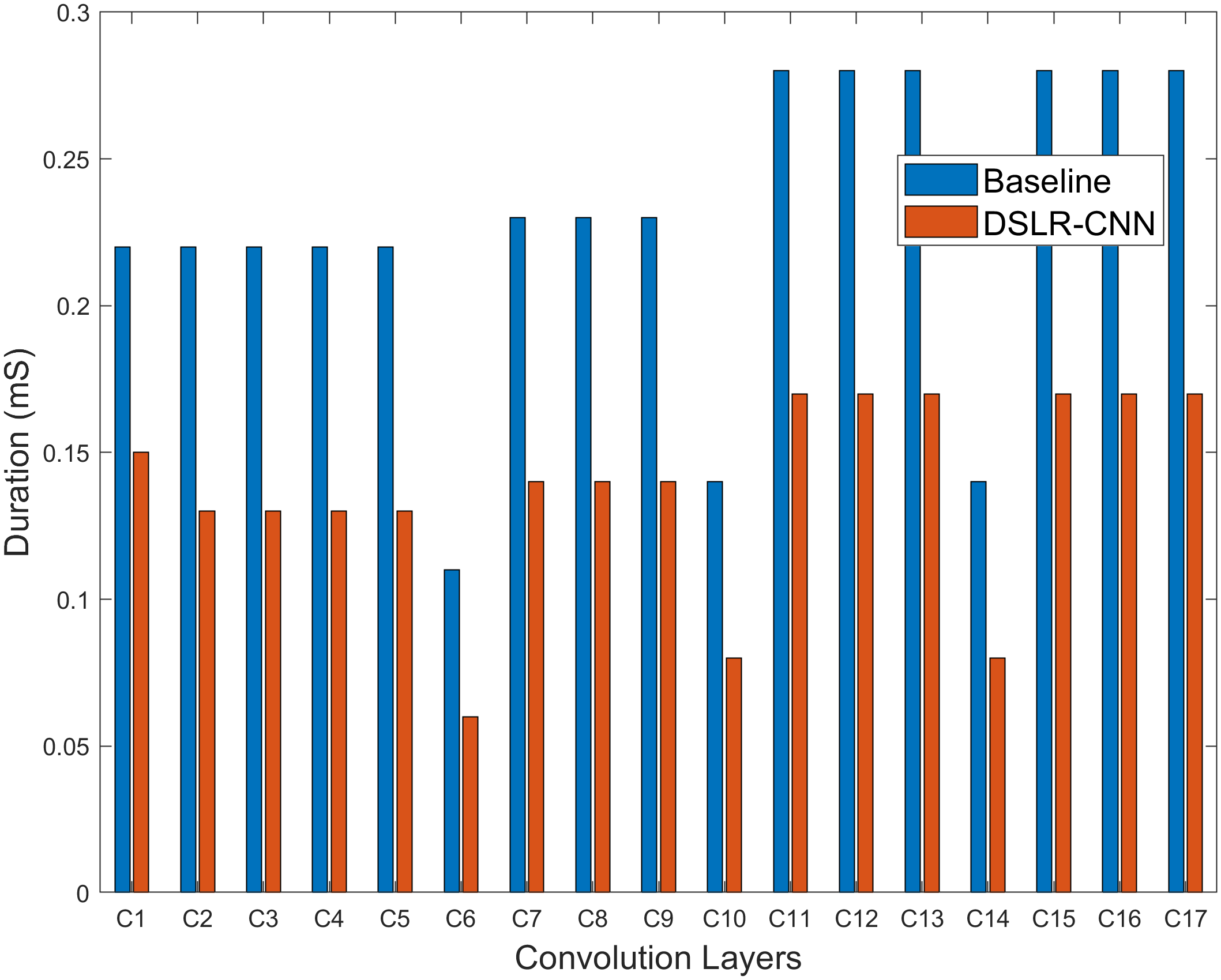}
        \caption{   }
    
    \end{subfigure}
    \vspace{0.5cm}  
    \begin{subfigure}[h]{1\linewidth}
        \centering
        \includegraphics[width=\linewidth]{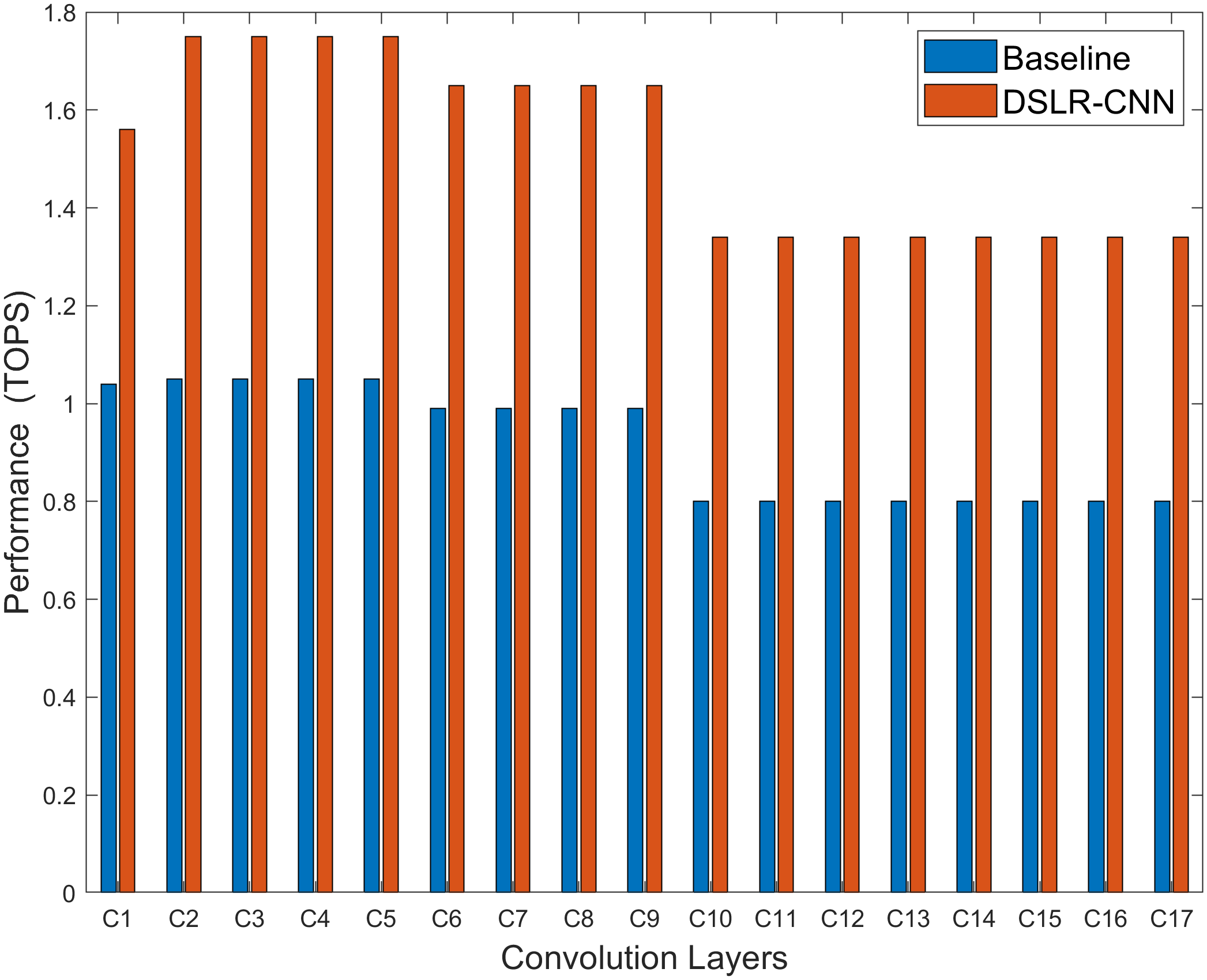}
        \caption{   }
        
    \end{subfigure}
    \caption{Comparison of ResNet Network: (a) Duration and (b) Performance.}
     \label{fig:resnetcomparison}
\end{figure}

\begin{figure}[ht]
	\begin{center}
 		\includegraphics*[width=\linewidth]{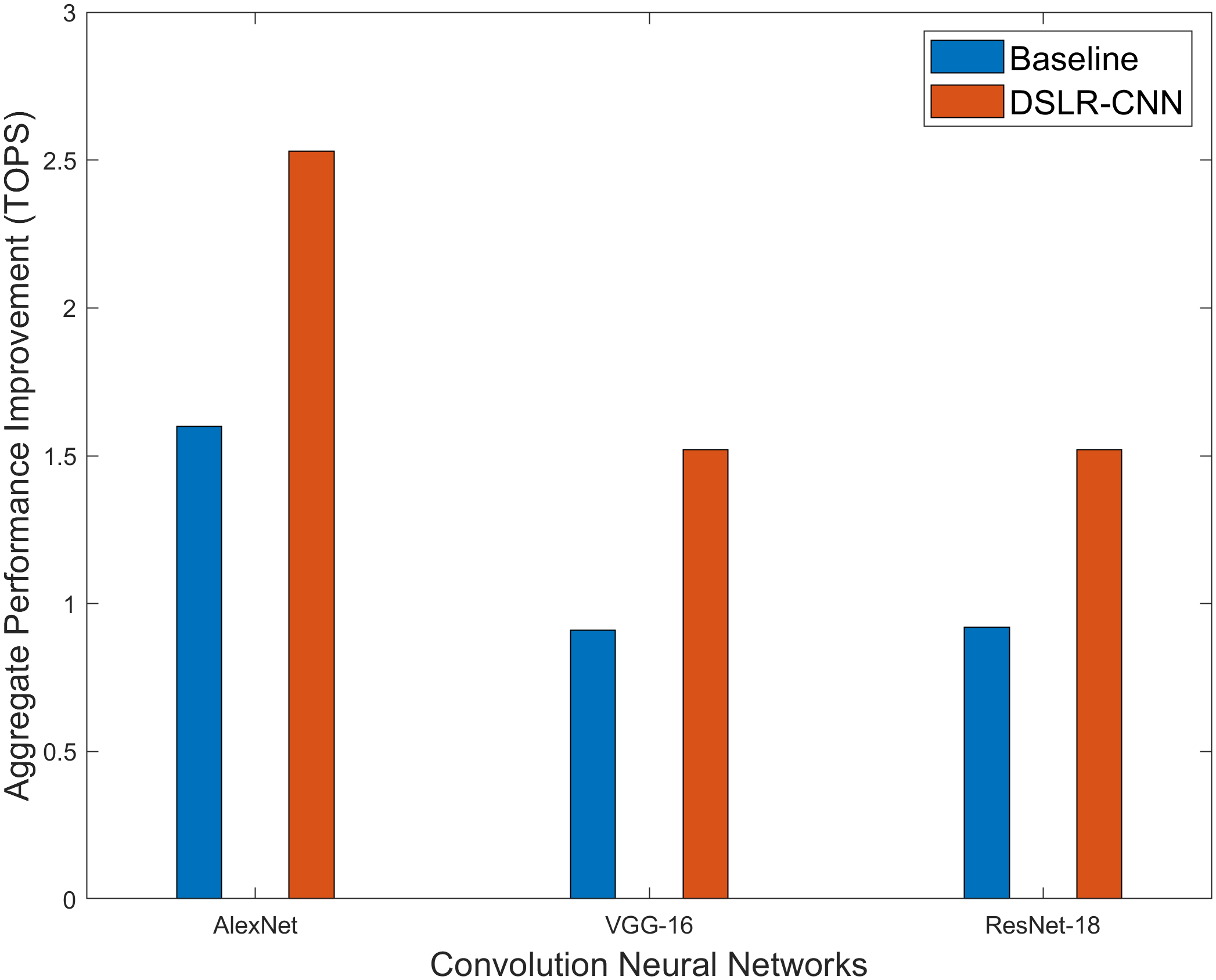}
 	\end{center}
 	\caption{Aggregate Performance Improvement of the DSLR-CNN Design Compared to the Baseline Across AlexNet, VGG-16, and ResNet-18 Networks.}
 	\label{fig: PI}
\end{figure}

Similarly, Fig.~\ref{fig:vggcomparison}, illustrates the inference time of the proposed DSLR-CNN design for the VGG-16 convolution layers, showcasing superior performance and reduced inference time compared to the baseline. Additionally, Fig.~\ref{fig:resnetcomparison} highlights the efficacy of the proposed DSLR-CNN design on the ResNet-18 network against the baseline, demonstrating superior performance and efficiency. Consistently, the proposed design exhibits remarkable performance across the AlexNet, VGG-16, and ResNet-18 networks. Aggregate performance enhancements are summarized in Fig.~\ref{fig: PI}, depicting significant improvements over the baseline design, across all evaluated convolutional neural networks. Specifically, the proposed design achieves a performance increase of $1.58 \times$ for AlexNet, $1.67 \times$ for VGG-16, and $1.65 \times$ for ResNet-18 network, as depicted in Fig.~\ref{fig: PI}.

\subsubsection{Operational Intensity of the proposed DSLR-CNN }

\begin{figure}[ht]
	\begin{center}
 		\includegraphics*[width=\linewidth]{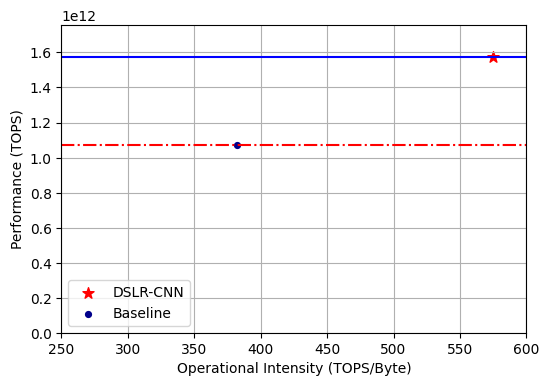}
 	\end{center}
 	\caption{Performance analysis of the DSLR-CNN design with the baseline design on first convolution layer of ResNet-18 Network in terms of performance (TOPS) and operational intensity (TOPS/Byte). }
 	\label{fig: roof}
\end{figure}

The implementation of an arithmetic-based accelerator design aims to significantly enhance performance and memory communication, as indicated by the operational intensity metric \cite{ofenbeck2014applying}. The effectiveness of this technique has been thoroughly analyzed by comparing it with the baseline design, as shown in Fig.~\ref{fig: roof}. The figure illustrates that the proposed DSLR-CNN design exhibits a higher operational intensity than the baseline design, highlighting the superior performance of the proposed approach. This finding suggests that the proposed technique outperforms the conventional bit-serial approach, especially when combined with the benefits of the LR arithmetic paradigm, enhancing operational intensity by $1.5\times$ compared to the baseline.

\section{Related Works}\label{sec:RW}

\subsection{Bit-Serial techniques of DNN} 

Several researchers have employed  bit-serial technique for the computation of convolution in CNNs based on the following observations: bit-serial arithmetic offers significant energy improvements,  it requires fewer resources compared to conventional bit-parallel computing, leading to reduced power consumption \cite{judd2015reduced}. By exploiting parallelism more efficiently and minimizing unnecessary data movement, bit-serial architectures can achieve higher energy efficiency, making them particularly attractive for resource- constrained environments. Moreover, bit-serial computing proves more efficient than conventional bit-parallel computing for CNNs as it can leverage the inherent parallelism within the networks, resulting in reduced memory usage, minimized interconnection requirements, and optimized bandwidth utilization. One such architecture, Stripes \cite{judd2016stripes}, uses bit-serial operating units to enhance performance and energy efficiency by dynamically adjusting precision to match the specific bit-width requirements of parameters and activations in each DNN layer. Also, Stripes is adaptable, allowing users to trade accuracy for further performance and energy improvements. However, replacing bit-serial architectures with adders or accumulators instead of conventional MAC units leads to longer latency in processing each weight and activation pair. The Stripes architecture is extended through the Unified Neural Processing Unit (UNPU) \cite{lee2018unpu}, which is a hybrid design that fixes the bit width of one operand while supporting variable bit widths for the other. This architecture incorporates lookup table (LUT)-based bit-serial processing elements (LBPEs) to accelerate DNN operations, significantly reducing the number of arithmetic operations and thereby improving both energy consumption and performance. To further enhance the flexibility of data transmission between LBPEs, the authors implemented a Network-on-Chip (NoC) interconnection. Neverthless, the extensive use of lookup tables results in substantial area overhead, making the UNPU design less efficient for implementing large-scale DNN models \cite{chen2021arbitrary}. Furthermore, various bit-serial architectures, including the bit-flexible design, known as Bit-Fusion \cite{sharma2018bit}, can dynamically fuse bit-level processing elements to match the bit-width of individual DNN layers. This flexibility in the architecture allows Bit Fusion to minimize the computation and communication at the finest granularity possible without losing accuracy. However, Bit-Fusion would incur a large area and energy overhead as it requires shift operations and heavy additions for reconfigurability \cite{inayat2023power}. Loom \cite{sharify2018loom}, exploits the use of mixed precision for both weights and activations, resulting in improved performance. However, Loom is a fully temporal design and can consume a significantly larger area and used more power \cite{sharma2018bit}. The array-based CNN accelerator, known as RASHT \cite{darbani2022rasht}, enhances resource utilization by dynamically resizing PEs to match the varying shapes of CNN layers. The core concept of RASHT is that a CNN network comprises layers of different sizes. Rather than employing a fixed PE engine for all layers, the engine adapts its size according to the specific layer it processes. This adaptive architecture results in significant improvements in both performance and energy efficiency.

Many architectures employing the loop-tiling strategy addressed the challenges of high computational complexity and excessive data storage in CNN hardware \cite{zhang2015optimizing}, \cite{sait2023optimization}. Some architectures utilize ring streaming dataflow and output reuse strategy to increase speed and reduce data access \cite{hsu2020essa}. The authors of Eyeriss \cite{chen2019eyeriss} introduced a spatial array architecture and a row-stationary dataflow \cite{chen2016eyeriss} to reduce data movement. By maximizing local data reuse, this approach enhances computational performance and reduces energy consumption. Conversely, Eyeriss faces limitations in data distribution, with about $50$\% of the total execution time spent on data transfer in the worst-case scenario \cite{darbani2022rasht}. To maximize hardware utilization and minimize data bandwidth, the design in \cite{lin2017data} introduced a run-time reconfigurable pipelined dataflow for CNN acceleration. By exploiting layer-specific characteristics and employing a tile-based model with an output-first strategy, this design significantly enhances area and bandwidth efficiency. The pipelined dataflow type is also flexible for various kernel sizes. Despite that, the approach leads to higher latency \cite{hsu2020essa}. To enhance inference performance, many designs aimed to utilize the zero-bits in neural networks and eliminate the corresponding operations induced by them. This approach helps compensate for the performance losses caused by bit-serial operations. The bit-serial designs in \cite{albericio2017bit}, \cite{delmas2019bit}, \cite{isobe2020low}, \cite{zhao2020bitpruner} have merely focused on exploiting the zero-bits in neural networks to eliminate the associated operations and improve inference performance. However, it is crucial to fully utilize the hardware units and minimize latency during computation cycles to avoid inefficiencies. To ensure accurate results and avoid errors, bit-serial computations require a specific order of operations, processing data from the least significant bit (LSB) to the MSB. Despite their efficiency, these designs often require complex nested loops over operand bits, contributing to increased computational complexity. Furthermore, structures relying on conventional arithmetic operators may significantly impact performance, power consumption, area utilization, and latency. Therefore, a promising solution to these challenges lies in unconventional arithmetic, featuring LR-based inner product computation units, which offer reduced energy utilization and require minimal response time, thus optimizing efficiency in CNN accelerators. 


\subsection{Comparison with the previous techniques}

To demonstrate the superiority of the proposed design, we compared various existing CNN accelerators, as outlined in Table~\ref{tab:compare}. In the case of the previous work, we relied on the reported results from their respective publications. As represented in the table, the proposed approach exhibits notable advantages regarding high performance, rapid response time, and exceptional energy efficiency. These benefits arise from the digit serial property inherent in LR arithmetic, which efficiently minimizes interconnects and signal activities through LR modules. Since prior accelerators exploit various techniques of neural networks in different frequencies, network types, and technology nodes,  making direct comparisons challenging, we adopted a scaling methodology for a more equitable comparison. Specifically, we scaled the synthesis results of the proposed DSLR-CNN design from 45nm to 65nm technology, to align with the previous studies, similar to the approach described in \cite{stillmaker2017scaling}. This adjustment allows us to provide a more accurate and fair comparison of performance and efficiency metrics. We have reported the highest peak performance of the proposed design in Table~\ref{tab:compare}. In terms of peak performance on the 45nm technology, the proposed design shows a significant performance boost of $14.92 \times$ and a $3.58 \times$ improvement in energy efficiency compared to DPNU \cite{shin201714}, which utilizes heterogeneous core architectures with specialized compute units for maximizing efficiency. When scaled to 65nm, the proposed design still maintains a strong advantage, delivering $10.62 \times$ higher performance and $1.57 \times$ better energy efficiency. Similarly, when compared with Eyeriss \cite{chen2016eyeriss}, a bit-parallel accelerator using a 16-bit fixed-point MAC with optimized row-stationary dataflow for enhanced DNN efficiency, the proposed design outperforms significantly, achieving $97.28 \times$ better performance and $18.84 \times$ higher energy efficiency on 45nm. After scaling, the proposed design achieves $69.24 \times$ better performance and $8.26 \times$ improved energy efficiency. The proposed design also surpasses the bit-serial neural network accelerator introduced in \cite{cheng2024leveraging}, which leverages dataflow techniques and architectural optimization. Specifically, the proposed design attains $569.11 \times$ better performance and $44.75 \times$ higher energy efficiency on 45nm, with the scaled results showing $405.10 \times$ and $19.62 \times$ improvements, respectively. Lastly, Bit-let \cite{lu2021distilling} and Bit-balance \cite{sun2024cim2pq} leverage bit-level sparsity and zero skipping to enhance performance and energy efficiency. However, the proposed design surpasses these models, achieving $12.02 \times$ and $4.37 \times$ better performance, respectively, while also delivering $13.76 \times$ and $3.80 \times$  superior energy efficiency on 45nm.  When scaled, the proposed design continues to outperform, showing $8.56 \times$ and $3.11 \times$ better performance, with energy efficiency improvements of $6.03 \times$ and $1.67 \times$.
.

\begin{table*}[ht]
\renewcommand{\arraystretch}{1.2}
\caption{ Overall Performance Comparison with Previous Works.}
\centering
\resizebox{1\textwidth}{!}{
\begin{tabular}{|c|c|c|c|c|c|c|c|c|c|}
\hline
Designs & DNPU \cite{shin201714} & Eyeriss \cite{chen2016eyeriss} & \cite{cheng2024leveraging} & Bit-let \cite{lu2021distilling} & Bit-balance \cite{sun2023bit} & DSLR-CNN & Scaled DSLR-CNN \\
\hline
Technology (nm) & 65 & 65 & 40 & 65 & 65 & 45 & 65 \\
\hline
Frequency (MHz) & 200 & 200 & 500 & 1000 & 1000 & 500&368 \\
\hline
Precision (Bits) & 16 & 16 & 8 & 16 & 16 & \multicolumn{2}{c|}{16} \\
\hline
Peak Performance (GOPS) & 300 & 46.04 & 7.87 & 372.35 & 1024 & 4478.97 & 3188.19 \\
\hline
Total Duration (ms) & - & 4309 & -& - & - & 0.94 & 1.28 \\
\hline
Power (mW) & 279 & 236 & 91.84 & 1390 & 1084 & 1249.42 & 2019.56 \\
\hline
Peak Energy Efficiency (TOPS/W) & 1.0 & 0.19 & 0.08 & 0.26 & 0.94 & 3.58 & 1.57 \\
\hline
\end{tabular}
}
\label{tab:compare}
\end{table*}


\subsection{Limitation and Challenges of the DSLR-CNN}

 LR arithmetic\cite{ercegovac2004digital}, which processes input operands serially and generates result digits from the LR, offers a superior computing paradigm that overcomes the limitations of conventional bit-serial designs. By processing data digit-by-digit, LR reduces latency and enhances parallelism, allowing multiple operations to be pipelined. This approach simplifies interconnections, leading to shorter signal paths and increased performance. Consequently, LR arithmetic is more energy efficient, scalable, and suitable for modern workloads such as deep learning, where real-time processing and high efficiency are critical. However, one of the limitations of LR arithmetic-based units is that they used distinct modules for multiplication and addition, resulting in an increased area. Another drawback is the inherent online delay, the fixed time required before the first digit of the result becomes available. This delay is a fundamental part of the online arithmetic process and contributes to increased system latency. Since the computation cannot produce results until after the online delay, this period is added to the total latency, which includes both the online delay and the time taken to process the remaining digits. In complex operations, particularly those involving multiple stages of arithmetic, these cumulative online delays can significantly impact total latency. In the future, we plan to explore the development of a composite online algorithm that integrates SoP directly within the same module, rather than using distinct online multiplication and addition units. By consolidating these operations, we aim to reduce overall area, power consumption, and latency. Instead of managing separate online delay modules, this approach will treat the latency as a unified process, potentially leading to significant performance improvements in efficiency and speed. To further demonstrate its effectiveness, we also intend to employ the application of LR arithmetic in various challenging deep learning areas such GoogleNet, YOLO, and transformers.

\section{Conclusion} \label{sec: conclusion}

This research aims to use low latency left-to-right bit-serial arithmetic-based sum-of-products units for convolution in CNN accelerators. The proposed DSLR-CNN design significantly outperforms the conventional bit-serial baseline design, demonstrating a $1.57\times$ performance improvement in AlexNet. For VGG-16 and ResNet-18, the design achieves greater performance improvements of $1.67\times$. Moreover, the design was synthesized using GSCL 45 nm technology, yielding remarkable results by surpassing conventional bit-serial design regarding latency. Furthermore, it improves operational intensity by $1.5 \times$ compared to conventional design. The DSLR-CNN technique achieves $1.05 \times$ to $1.06 \times$ more $TOPS/Watt$ metrics outperforming the baseline design by showing superior peak energy efficiency when evaluated for the convolution layer of AlexNet, VGG-16, and ResNet-18. The comparison with the previous techniques illustrates the efficacy of the proposed approach in terms of peak performance and peak energy efficiency. In future, we aim to extend the scope of this research to accelerate more modern and complex DNN architectures.

\bibliographystyle{ieeetr}
\bibliography{Reference}

\EOD

\end{document}